\documentclass[usenatbib,usegraphicx]{mn2e}
\usepackage{times}

\newcommand{\vvir}{\ensuremath{V_{\mathrm{vir}}}}
\newcommand{\msun}{\ensuremath{\mathrm{M_{\odot}}}}
\newcommand{\kms}{\ensuremath{\mathrm{km\ s^{-1}}}}
\newcommand{\fe}{\ensuremath{\langle\mathrm{Fe}\rangle}}
\newcommand{\mgb}{\ensuremath{\mathrm{Mg}\,b}}
\newcommand{\mgfe}{\ensuremath{[\mathrm{MgFe}]}}
\newcommand{\hbeta}{\ensuremath{\mathrm{H}\beta}}

\newcommand{\logt}{\ensuremath{\log t}}
\newcommand{\tssp}{\ensuremath{t_{\mathrm{SSP}}}}
\newcommand{\z}{\ensuremath{\mathrm{[Z/H]}}}
\newcommand{\zssp}{\ensuremath{\mathrm{[Z/H]_{SSP}}}}

\newcommand{\enh}{\ensuremath{\mathrm{[E/Fe]}}}

\begin{document}

\pubyear{2009}
\pagerange{1--17}

\title[Probing recent star formation]{Probing recent star formation
  with absorption-line strengths in hierarchical models and
  observations}
 
\author[S.~C. Trager \& R.~S. Somerville]{S. C. Trager$^1$\thanks{email:
  sctrager@astro.rug.nl} and R. S. Somerville$^{2,3}$\\
  $^1$Kapteyn Astronomical Institute, University of Groningen, Postbus
  800, NL-9700 AV Groningen, The Netherlands\\ 
  $^2$Max-Planck-Institut f\"ur Astronomie, K\"onigstuhl 17, D-69117
  Heidelberg, Germany\\
  $^3$Space Telescope Science Institute, 3700 San Martin Drive,
  Baltimore, MD 21218, USA}

\maketitle

\begin{abstract}
Stellar population parameters (e.g. age, metallicity, and stellar
abundance ratios) derived from spectral line-strengths provide a
powerful probe of galaxy properties and formation histories. We
implement the machinery for extracting line-strengths and
`single-stellar-population-equivalent' (SSP-equivalent) stellar
population parameters from synthetic spectra generated by a
hierarchical galaxy formation model. Our goals are (1) to test the
consistency of these line-strength-derived stellar population
parameters with more physically relevant light- and mass-weighted
parameters for complex, cosmologically-motivated star-formation
histories, (2) to interpret line-strength observations for early-type
galaxies within the context of hierarchical structure formation, and
(3) to test the galaxy formation models using stellar population
parameters derived from some of the best available samples of observed
line-strengths.  We find that the SSP-equivalent age is related to the
light-weighted age in a complicated fashion that reflects the
influence of recently-formed stars and is poorly correlated with the
mass-weighted age.  We find that the tendency for SSP-equivalent ages
to be biased young means that `archaeological downsizing' overstates
the `true', mass-weighted downsizing in age with mass.  We find
however that the SSP-equivalent metallicity closely tracks the mass-
and light-weighted metallicities, so that observed mass--metallicity
relations for old galaxies closely reflect the underlying trends.  We
construct mock catalogues of early-type galaxies in a Coma
cluster-sized halo and compare them directly to observations of
early-type galaxies in the Coma cluster.  The similarity of the
SSP-equivalent ages in the observational samples and the mock
catalogues gives us confidence that the star-formation quenching
implemented in the hierarchical galaxy formation model produces
roughly the correct amount of recent star formation.  Unfortunately,
the current observational samples are either too small or have too low
signal-to-noise to accurately determine detailed star-formation
histories.  However, the data show that the model has
deficiencies: the SSP-equivalent metallicities are too low and have
the wrong slope as a function of velocity dispersion, and the
SSP-equivalent ages of the model galaxies may have an incorrect slope
as a function of velocity dispersion.  These problems are indicative
both of the simplified chemical evolution prescription currently
implemented in the galaxy formation model and that the star-formation
histories resulting from the model are incorrect in detail.
\end{abstract}


\section{Introduction}
\label{sec:introduction}

The optical colours of old stellar populations are determined by an
interplay between the colours of the red giant branch and the
main-sequence turn-off, which together produce nearly all of the light
at these wavelengths.  This interplay results in the well-known
age--metallicity degeneracy in old stellar populations
\citep[e.g.,][]{OConnell86}: a change by a factor of two in age looks
the same in optical colours (and metal-line strengths) as a factor of
three change in metallicity \citep{W94}.  This age--metallicity
degeneracy long limited -- and to some extent, still limits -- our
understanding of the stellar populations of `red, dead' galaxies like
early-type galaxies.

In a pioneering paper, \citet{Rabin82} demonstrated that it is
possible to break the age--metallicity degeneracy by using the
strength of the Balmer lines of hydrogen as a function of a metal
line.  Ever since this breakthrough, the potential power of `fossil'
star formation evidence in local galaxies has been expected to yield
insights on the formation of these objects complementary to direct
look-back studies \citep[see, e.g., the recent review
  by][]{Renzini06}.  If we could measure the mass-weighted ages of
local galaxies, we would know when the bulk of their stars were
formed, a key constraint on galaxy formation models.  This approach
has yielded very interesting results when authors have applied modern
stellar population models \citep[by, e.g.,][and
  others]{Buzzoni94,W94,Vazdekis96,BC03,TMB03,Schiavon07} to
high-quality spectral observations of early-type galaxies. For
example, these studies have demonstrated the wide range of galaxy ages
\citep{G93}, the dependence of stellar population parameters (age,
metallicity, and abundance ratio) on galaxy velocity dispersion
\citep{T00b}, the dependence of age and star-formation time-scale on
mass (often referred to as `fossil' or `archaeological downsizing':
\citealt{TMBO05,Nelan05}), and the dependence of galaxy age and star
formation time-scale on environment \citep[but see contradicatory
  evidence in \citealt{Thomas07}]{TMBO05,Bernardi06,SB06b}.

But as age-dating techniques become more routine and confront larger
data sets, the pitfalls of this approach become more apparent.  The
most dangerous pitfall is that the easiest and therefore most common
application of these techniques assumes that galaxies can be
meaningfully parametrized by a single `age'.  This assumption is in
clear violation of the modern picture of galaxy formation, in which
(early-type) galaxies do not form in a single burst, but are built up
through a sequence of mergers \citep[e.g.,][among many
  others]{Toomre77,WR78,BFPR84,KWG93,Cole94,Kauffmann96,KC98,deLucia06}.
As shown schematically by \citet{T00b} and quantitatively by
\citet{ST07}, the addition of a small fraction of young stars to an
old population strongly biases the \emph{apparent} age of a galaxy.
As a simple example, the addition of 2 per cent by mass of 1 Gyr-old
stars to a 12 Gyr-old population results in an apparent age of 5 Gyr
\citep{TFD08}.  The extreme sensitivity of stellar population ages to
recent star formation leads to ambuigities in interpretation: is a
galaxy truly young or just composite, as would be expected from our
current understanding of galaxy formation?\footnote{Although more
  sophisticated methods of disentangling star-formation histories from
  spectra are becoming available
  \citep[e.g.,][]{MOPED1,MOPED2,Ocvirk06a,VESPA}, they appear either
  to require very high signal-to-noise and high-resolution spectra
  over a long spectral baseline \citep{Ocvirk06a,VESPA} or to produce
  results that are useful on average only for very large samples
  \citep[e.g.][]{Panter03,Jimenez07}.}

Furthermore, accurate age dating, even assuming a single stellar
population, requires very accurate line strengths and therefore very
high signal-to-noise spectra
\citep[$\mathrm{S/N}>100/\mathrm{\AA}$:][]{T97,Cardiel98,Cardiel03,Kuntschner01}.
This limits its usefulness so far either to small data sets
\citep[e.g.,][]{G93,Kuntschner00,TMBO05,SB06a,TFD08} or to the need to
combine dozens-to-hundreds of galaxies together, thus losing
information about dispersions in stellar population properties
\citep[e.g.,][]{Bernardi05,Graves07}.  It is therefore clear that
guidance from simulations could be very useful for understanding the
impact of both complex star-formation histories and noisy data on the
interpretation of stellar population ages, \emph{if simulation results
  could be analyzed, to the greatest extent possible, in exactly the
  same way as the observations}.

We turn to the latest generation of hierarchical galaxy formation
models to provide this guidance.  This new generation of models
implements feedback from active galaxy nuclei, most importantly in the
form of heating from radio jets, to solve the overcooling
\citep[e.g.,][]{KWG93} and star-formation quenching
\citep[e.g.,][]{somerville:04} problems of the previous generation of
models \citep{Bower06,Cattaneo06,Croton06,s08}.  These models
successfully predict many of the properties of local galaxies, in
particular their luminosity functions, the dichotomy of galaxy colours
\citep[e.g.,][]{Strateva01}, the morphology of the colour--magnitude
diagram, and even its evolution to $z=1$.  However, the authors of
these papers make predictions almost solely in colour--magnitude
space, which is at best a blunt tool: as discussed above, the
age--metallicity degeneracy makes interpreting the colours of old
populations problematic.

Now that the hierarchical galaxy formation models are doing a
reasonable job at predicting basic local galaxy properties, we
investigate using these models to produce `mock catalogues' to guide
our interpretation of line-strength observations of local galaxies.
We further want to subject the models themselves to a sharper test
than the blunt tool of colours alone.  We have two desires in this
paper to make these goals concrete: (1) to use a well-defined sample
of galaxies drawn from a well-understood environment that can be
simply modelled in the context of the hierarchical galaxy formation
models and (2) to bring the models as close to the `observational
plane' as possible.  To meet the first desire, we use line-strength
observations from recent samples of early-type galaxies in the Coma
cluster.  We choose the Coma cluster because it is a rich,
well-studied cluster whose dark matter halo properties are relatively
well-understood \citep{LM03,Kubo07}.  The Coma cluster is close enough
to allow detailed spectroscopic study of a large number of early-type
galaxies and yet is far enough that the galaxies can be treated almost
as `point sources' (but not quite: see the Appendix) in a single dark
matter halo, a convenience when comparing to the hierarchical models.
We will find below that the current line-strength observation samples
of Coma cluster galaxies do not provide the ideal galaxy sample for
our needs, but they are the best we have at the moment.  To meet the
second desire, we create synthetic spectra from the hierarchical
galaxy formation models, based on the predicted star formation and
chemical enrichment histories convolved with stellar population
models, and run them through the same machinery used to extract
line-strengths, SSP-equivalent parameters, and error-bars on these
parameters from the observed spectra. We apply this machinery to mock
catalogues drawn from simulated Coma-sized halos such that we
reproduce the Velocity Dispersion Function (VDF) of a particular
observational sample.  These predictions give us the sharp tool we
need to understand the properties of local galaxies and to test the
models themselves.

We have three goals in the current paper: (1) understanding and
calibrating how SSP-equivalent parameters extracted from spectral
line-widths represent the underlying mass- and luminosity-weighted
ages and metallicities for galaxies with complex, cosmologically
motivated star formation and enrichment histories, (2) interpreting
line-strength observations of early-type galaxies within the context
of hierarchical models, and (3) using the observed line-strengths for
the best available samples of Coma cluster galaxies to test the
physics in the galaxy formation models. The paper is organized as
follows.  In Section~\ref{sec:models} we review the ingredients of the
hierarchical galaxy formation models, their calibration, and the
stellar population modelling.  In Section~\ref{sec:popresults} we
examine the SSP properties of composite populations from synthetic
spectra created via the hierarchical galaxy formation models.  In
Section~\ref{sec:coma} we compare models of galaxies in the Coma
cluster with observations and suggest improvements for both models and
observations.  Finally in Section~\ref{sec:conclusions} we summarise
our findings.  An appendix details the aperture corrections needed to
compare the resolved galaxy observations to the unresolved model
galaxies.

\section{Hierarchical galaxy formation models}
\label{sec:models}

We make use of a semi-analytic galaxy formation model described in
detail in \citet{sp}, \citet{spf} and \citet[][hereafter S08]{s08}.

\subsection{Merger trees and substructure}
\label{sec:mergertrees}

The models are based on Monte Carlo realizations of dark matter halo
merger histories, constructed using a slightly modified version of the
method of \citet{sk:99}, as described in S08. The merging of sub-halos
within virialized dark matter halos is modelled by computing the time
required for the sub-halo to lose its angular momentum via dynamical
friction and fall to the center of the parent halo. Our treatment
takes into account the mass loss of the orbiting dark matter halos due
to tidal stripping, as well as tidal destruction of satellite
galaxies.

\subsection{Gas cooling and star formation}
\label{sec:sf}

When gas cools, it is assumed to initially settle into a thin
exponential disc, supported by its angular momentum. Given the halo's
concentration parameter, spin parameter, and the fraction of baryons
in the disc, we can use angular momentum conservation arguments to
compute the scale radius of the exponential disc after collapse
\citep{somerville:08b}.

The models make use of fairly standard recipes for the suppression of
gas infall due to a photo-ionizing background, cooling, star
formation, and supernova feedback, as described in S08. Quiescent star
formation, in isolated discs, is modelled using the empirical
Kennicutt star-formation law \citep{kennicutt:89,kennicutt:98}:
\begin{equation}
\dot{\Sigma}_{\mathrm{SFR}} = A_{\mathrm{Kenn}} \,
{\Sigma_{\mathrm{gas}}}^{N_K} ,
\end{equation}
with $A_{\mathrm{Kenn}} = 8.33\times10^{-5}$ (a factor of two lower
than the canonical value, see S08), $N_K=1.4$, $\Sigma_{\mathrm{gas}}$
the surface density of cold gas in the disc (in units of $\msun\,
\mathrm{pc}^{-2}$), and where $\Sigma_{\mathrm{SFR}}$ has units of
$\msun\,\mathrm{yr}^{-1}\,\mathrm{kpc}^{-2}$.  We also adopt a
critical surface density threshold $\Sigma_{\mathrm{crit}}$, and
assume that only gas lying at surface densities above this value is
available for star formation.

\subsection{Galaxy mergers and morphologies}
\label{sec:mergers}

Galaxy mergers may induce a burst of star formation and destroy any
pre-existing disc component, depending on the mass ratio of the
smaller to the larger galaxy, $\mu$. We parameterize the strength and
timescale of these bursts according to the results of a large suite of
hydrodynamic simulations of galaxy mergers \citep{robertson:06,cox:08}
as described in S08. The burst efficiency is a function of the mass
ratio $\mu$ and the bulge-to-total fraction of the primary galaxy
(larger mass ratios yield more dramatic bursts, while the presence of
a large bulge further suppresses bursts in low mass-ratio mergers);
the burst timescale is primarily a function of the primary progenitor
galaxy's circular velocity (lower circular velocity galaxies produce
longer bursts).

Mergers can also heat and thicken, or even destroy, a pre-existing
disc component, driving galaxies towards morphologically earlier
Hubble types. We assume that the fraction of the pre-existing stars
that is transferred from a disc to a spheroidal component is a
strongly increasing function of the merger mass ratio $\mu$, such that
minor mergers with $\mu < 0.2$ have little effect, and major mergers
with $\mu > 0.25$ leave behind a spheroid-dominated remnant.

Morphologies of the model galaxies are determined using the
bulge-to-total $B$-band light ratios $B/T$ following the scheme
described in \citet{KWG93} and \citet{sp}, based on the results of
\citet{SdV86}.  Galaxies with $B/T<0.405$ are considered to be
spirals, galaxies with $0.405<B/T<0.603$ are considered to be S0s, and
galaxies with with $B/T>0.603$ are considered to be ellipticals.  In
the current paper we consider galaxies with type S0 or E to be
early-type galaxies.  Note that we conservatively use $B$ magnitudes
\emph{uncorrected} for dust extinction in this calculation.  This
choice moves galaxies to somewhat later morphological types (smaller
$B/T$) as the discs are extinguished more than the bulges, so the
unextinguished discs are brighter than they would be observed.

It is important to note here that although the semi-analytic model
provides predictions of size estimates for disc galaxies (see S08),
the model currently does not provide size estimates for bulges.
Therefore when comparing properties of local bulge-dominated
(early-type) galaxies to model predictions, we must be aware of the
possible effects of spatial gradients on these results.  We discuss
this in more detail in Sec.~\ref{sec:coma} and in the Appendix.

\subsection{Black hole growth, supernovae, and feedback}
\label{sec:feedback}

In our model, mergers also trigger the accretion of gas onto
supermassive black holes in galactic nuclei. Each top-level halo in
our merger trees is seeded with a black hole of mass $M_{\mathrm{seed}}
\simeq 100 \msun$. Following a merger, the black hole is allowed to
grow until the radiative energy being emitted by the AGN becomes
sufficient to halt further accretion. This self-regulated treatment of
black hole growth is based on hydrodynamic simulations including BH
growth and feedback \citep{springel:05a,dimatteo:05,hopkins_bhfpth:07}
and is described in more detail in S08. Because the potential well
depth of the spheroid determines how large the black hole has to grow
in order to halt further accretion, this model successfully reproduces
the black hole mass vs. spheroid mass relation at $z\sim0$. Energy
radiated by black holes during this `bright', quasar-like mode can
also drive galactic-scale winds, clearing cold gas from the
post-merger remnants (see S08).

In addition to the rapid growth of BH in the merger-fueled,
radiatively efficient `bright mode', we assume that BH also
experience a low-Eddington-ratio, radiatively inefficient mode of
growth associated with efficient production of radio jets that can
heat gas in a quasi-hydrostatic hot halo. The accretion rate in this
phase is modelled assuming Bondi accretion using the isothermal
cooling flow solution of \citet{nulsen_fabian:00}. We then assume that
the energy that effectively couples to and heats the hot gas is given
by $L_{\mathrm{heat}} = \kappa_{\mathrm{heat}} \eta
\dot{m}_{\mathrm{radio}} c^2$, where $\dot{m}_{\mathrm{radio}}$ is the
accretion rate onto the BH, $\eta=0.1$ is the assumed conversion
efficiency of rest-mass into energy, and $\kappa_{\mathrm{heat}}$ is a
free parameter of order unity.

Cold gas may be reheated and ejected from the galaxy, and possibly
from the halo, by supernova feedback. The rate of reheating of cold
gas is given by
\begin{equation}
\dot{m}_{\mathrm{rh}} = \epsilon^{SN}_0 \left( \frac{200\,\kms}{V_c}
\right)^{\alpha_{\mathrm{rh}}}\dot{m}_*,
\end{equation}
where $\alpha_{\mathrm{rh}}=2$ is assumed for energy-driven winds
(S08)\footnote{Note that there is a typographical error in this
  equation in S08. The expression given here is the correct one.}.
The heated gas is either trapped within the potential well of the dark
matter halo and deposited in the `hot gas' reservoir, or is ejected
from the halo into the `diffuse' Intergalactic Medium (IGM). The
fraction of reheated gas that is ejected from the halo is given by:
\begin{equation}
f_{\mathrm{eject}}(\vvir) =
\left[1.0+(\vvir/V_{\mathrm{eject}})^{\alpha_{\mathrm{eject}}}
  \right]^{-1},
\end{equation}
where $\alpha_{\mathrm{eject}}=6$ and $V_{\mathrm{eject}}$ is a free
parameter in the range $\simeq 100$ --$150\,\kms$. This ejected gas is
reincorporated into the halo on a timescale proportional to the halo
dynamical timescale (see S08).

\subsection{Chemical evolution}
\label{sec:gce}

Each generation of stars produces new heavy elements. The mass of
metals produced is $dM_Z = y\, dm_*$, where $dm_*$ is the mass of
stars produced and $y$ is the yield. New metals are added directly to
the cold gas. When the cold gas is reheated or ejected by supernova
feedback, the metals are transferred to the hot or ejected component,
respectively, in the same manner as the gas. Ejected metals are also
`re-accreted' by the halos as described in S08. Note that the simple
chemical enrichment model used here includes only enrichment by Type
II supernovae. Therefore the metallicities quoted should be
interpreted as representing $\alpha$-type elements.  In this study we
assume that $\enh=0$, that is, that the galaxies have solar abundance
ratios, and that the metallicities quoted are total metallicities.
This assumption has some impact on the results discussed below, in
particular on the inferred metallicities of the model galaxies.  In a
forthcoming paper (Arrigoni et al., in prep.) we relax this assumption
and include a full galactic chemical evolution (GCE) model, tracing up
to 19 individual elements.  We point out below where this full GCE
model differs from the present models.

\subsection{Calibration}
\label{sec:calibration}

The models contain a number of free parameters, which are summarized
in Table 1 of S08. We adopt the `WMAP3' \citep{WMAP3} values of the
cosmological parameters (S08).  We then set the free parameters by
matching certain observations of nearby galaxies, as described in
detail in S08. For example, the star formation parameters
$A_{\mathrm{Kenn}}$ and $\Sigma_{\mathrm{crit}}$ are constrained by
the observed gas fractions of nearby spiral galaxies. The supernova
feedback parameters are set in order to reproduce the stellar mass
fractions in low-mass halos ($\la10^{12}\msun$) such that the low-mass
or low-luminosity slope of the observed stellar mass or luminosity
function is reproduced. The yield $y$ is set such that spiral galaxies
in Milky Way sized halos ($2\times10^{12}\msun$) have approximately
solar metallicity. The parameter controlling radio mode AGN feedback
is set to give the minimum amount of feedback required to solve the
overcooling problem and bring the high-mass or high-luminosity end of
the galaxy stellar mass or luminosity function into agreement with
observations. The resulting models, which are tuned solely using
$z\sim0$ observations, are also in fairly good agreement with direct
probes of the mass assembly and star-formation history of the universe
from look-back studies based on deep cosmological surveys (S08).

\subsection{Stellar kinematics}
\label{sec:sigma}

To compare model galaxy properties with observations, a useful
observational tracer is the one-dimensional line-of-sight velocity
dispersion, $\sigma$ (hereafter referred to simply as the velocity
dispersion).  Velocity dispersion appears to be the controlling
parameter of stellar population observables, including colour and
luminosity \citep{Bernardi05}, metallicity and abundance ratios
\citep{T00b} and possibly even age \citep[e.g.][]{TMBO05,Nelan05}.

Unfortunately we do not currently have detailed predictions of
velocity dispersions available for our model galaxies. However, it is
well known that in the local universe, there is a tight correlation
between stellar mass and velocity dispersion for early type
galaxies. We therefore simply use an empirical conversion,
\begin{equation}
  \log\sigma = 2.2 - 0.1023\left[21.49 - \left(\frac{\log M_* -
      1.584}{0.433}\right)\right]. \label{eq:sigma}
\end{equation}
This relation is the result of fitting to the stellar mass and
velocity dispersion data of \citet{HR04} and is consistent with the
results of \citet{LFL06} obtained from SDSS. This parametrisation of
$\sigma$ is a better approximation to the stellar velocity dispersion
than estimates that simply assume that it is proportional to the
velocity dispersion of the dark matter (sub)halo within the virial
radius \citep[e.g.][]{springel:01}. Early-type galaxies are baryon
dominated so the stellar velocity dispersion can scale quite
differently from the (sub)halo velocity dispersion.

\subsection{Creating synthetic spectra}
\label{sec:stellarpops}

We convolve the resulting star-formation histories (SFHs) of the model
galaxies with the multi-metallicity stellar SED models of
\citet[hereafter BC03]{BC03}, based on the Padova1994 \citep{Padova}
isochrones with a \citet{Chabrier01} initial mass function (IMF), to
produce synthetic spectra, broad-band magnitudes and colours,
absorption-line strengths, and mass-to-light ratios.

The predicted magnitudes, line-strength indices, mass-to-light ratios,
light-weighted stellar population parameters, and inferred
SSP-equivalent stellar population parameters of the galaxy formation
models are computed in a `post-processing' step by combining the SFHs
with the stellar population models.  In detail, the SFHs provide a
mass in stars formed at each point in a grid of ages ($t$) and
metallicities (\z) for each galaxy in the model at the redshift of
interest (in this paper, $z=0.023$; see below).  The magnitudes in
photometric band $k$ are computed as
\begin{equation}
  M_k=-2.5\log\left(\sum_i\sum_j {\cal M}_{i,j}I^k_{i,j}\right) +
  \mathrm{ZP}_k
\end{equation}
where $i,j$ are the age and metallicity bins, ${\cal M}_{i,j}$ is the
mass formed at age $i$ and metallicity $j$, $I^k_{i,j}$ is the
intensity \emph{per unit mass} in band $k$ for the stellar population
model with age $i$ and metallicity $j$, and $\mathrm{ZP}_k$ is the
magnitude zero-point for band $k$.  We compute both dust-extinguished
and unextinguished magnitudes.  We model the impact of dust on the
magnitudes using an approach similar to that proposed by
\citet{charlot_fall:00} as modified by \citet{delucia_blaizot:07}. We
include two dust components, one due to `cirrus' in the disc and one
associated with the dense `birth clouds' surrounding young star
forming regions. The face-on optical depth of the `cirrus' dust in the
V-band is given by
\begin{equation}
\tau_{V,0} \propto \tau_{\mathrm{dust},0}\, Z_{\mathrm{cold}}
m_{\mathrm{cold}}/(r_{\mathrm{gas}})^2,
\end{equation}
where $\tau_{\mathrm{dust},0}$ is a free parameter,
$Z_{\mathrm{cold}}$ is the metallicity of the cold gas,
$m_{\mathrm{cold}}$ is the mass of cold gas in the disc, and
$r_{\mathrm{gas}}$ is the radius of cold gas in the disc, which we
assume to be 1.5 times the scale radius of the stars. We then use a
`slab' model to compute the inclination-dependent extinction
\citep[see][]{sp}. Young stars ($t < 10^7$ yr) are additionally
enshrouded in a screen of dust with optical depth $\tau_{\mathrm{BC},
V} = \mu_{\mathrm{BC}} \tau_{V,0}$. We take $\mu_{\mathrm{BC}}=3$. We
assume a Galactic extinction curve \citep{cardelli:89} for the cirrus
component and a power-law extinction curve
$A_{\lambda}\propto(\lambda/5500\,\mathrm{\AA})^n$, with $n=0.7$, for
the birth clouds \citep{charlot_fall:00}.

Line-strength indices are computed similarly to the magnitudes as
\begin{equation}
  \mathrm{EW} = \Delta \left(1-\frac{\sum_i\sum_j {\cal
      M}_{i,j}F_L^{i,j}}{\sum_i\sum_j {\cal M}_{i,j}F_C^{i,j}}\right),
\end{equation}
where $\Delta$ is the width of the index in \AA, $F_L^{i,j}$ is the
flux in the line centre for the stellar population model with age $i$
and metallicity $j$ and $F_C^{i,j}$ is the flux in the continuum for
that model, for indices measured in \AA, and as
\begin{equation}
  \mathrm{Mag} = -2.5\log\left(\frac{\sum_i\sum_j {\cal
      M}_{i,j}F_L^{i,j}}{\sum_i\sum_j {\cal M}_{i,j}F_C^{i,j}}\right) 
\end{equation}
for indices measured in magnitudes \citep{W94}.  We make no attempt to
account for dust in the line-strength indices \citep{MacArthur05}.  

\subsection{Determining SSP-equivalent parameters}
  
As discussed in the Introduction, the most common approach for
interpreting observed spectral line-strengths in terms of physical
parameters that probe galaxy formation is to assume that the stellar
population in a galaxy is a `simple stellar population', i.e., that it
is comprised of stars of a single age and metallicity.  Under this
assumption, one can use stellar population models to derive the
SSP-equivalent age and metallicity that fits the observed
line-strengths.



To break the age--metallicity degeneracy, we must use \emph{both} a
metal line and a hydrogen line.  For the analysis herein, we use the
metal-line indices \mgb, Fe5270, and Fe5335, which are more sensitive
to metallicity than to age, and a Balmer-line index, which is more
sensitive to age than metallicity \citep{W94,WO97}.  Throughout the
present paper we use the \hbeta\ index, as this is nearly always
combined with the previously mentioned metal-line indices to determine
SSP-equivalent parameters \citep[e.g.][]{T00a,TMBO05}.  Using a
variety of other age and metallicity indicators is beyond the scope of
this paper.

We determine stellar population parameters directly using a non-linear
least-squares fitting code based on the Levenberg-Marquardt algorithm
described in \citet[hereafter TFD08]{TFD08}, in which stellar
population models are linearly interpolated in (\logt, \z,
\enh)\footnote{Recall that the galaxy formation models have solar
  abundance ratios and therefore $\enh=0$, although this is not true
  for the galaxy observations described later.} on the fly to produce
line strength indices (and magnitudes, colours and other parameters,
but these are not typically used in the fitting process).  For this
purpose we use the BC03 stellar population models modified as
described in TFD08.  If uncertainties in the stellar population
parameters are desired, they are computed by taking the dispersion of
stellar population parameters from 10000 Monte Carlo trials using the
provided index errors as 1-$\sigma$ Gaussian deviates. We compute the
stellar population quantities for both the entire galaxy (`total') and
its bulge component (`bulge').  Finally, we compute light-weighted
stellar population parameters for each photometric band by combining
the star formation histories with the SEDs.

\subsection{Coma Cluster model}

For the analysis in the current paper we are interested in comparing
our models with real observations of early-type galaxies.  The Coma
cluster is a convenient target, as a large body of
reasonable-to-excellent quality line-strength data is available (e.g.,
see the compilation in TFD08).  We therefore generated twenty
realizations of a Coma cluster-sized halo, i.e., a halo with a
circular velocity $V_c=1066\,\kms$, corresponding to a virial mass of
$M_{\mathrm{vir}}=10^{14.9}\,\msun$, similar to if slightly lower than
recent estimates of the mass of the Coma cluster \citep[and references
  therein]{LM03,Kubo07}.  The smallest halos that we resolve in our
merger trees have a virial mass of $10^{10} \msun$, so we select
galaxies with $M_*>10^9\,\msun$, corresponding to a velocity
dispersion $\sigma=57\,\kms$ using Eq.~\ref{eq:sigma}.  We select only
`early-type' galaxies (E and S0) according to their B-band
bulge-to-total light ratios, as described in
Section~\ref{sec:mergers}.  We set the output redshift of the
simulations to be $z=0.023$ \citep[e.g.][]{Hudson01} to match the
redshift of the Coma cluster.

\section{Single-stellar-population properties of galaxies in
  hierarichal models}
\label{sec:popresults}

In this section we seek to understand the properties of composite
stellar populations when interpreted using single-stellar-population
(SSP) equivalent parameters.  This can be thought of as extending the
work of \citet{ST07} to more complicated -- and hopefully more
realistic -- star-formation histories.  We first examine the SFHs and
metallicity distributions of galaxies drawn from semi-analytic models,
then examine their absorption-line strengths and inferred
SSP-equivalent ages and compare these with the intrinsic physical
timescales predicted by the semi-analytic models, such as luminosity-
and mass-weighted ages, merger times, and birthrates. 

\subsection{Star-formation histories and metallicity distributions}
\label{sec:tzhists}

\begin{figure}
  \includegraphics[width=85mm]{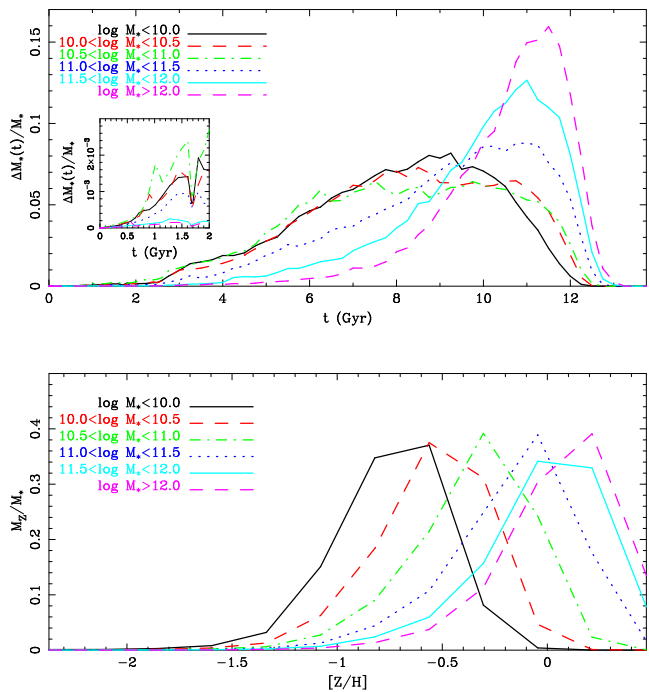}
  \caption{Average star-formation histories (top) and metallicity
    distributions (bottom) for early-type galaxies in 20 realizations
    of a Coma cluster-sized halo, binned by stellar mass. The inset in
    the top panel shows the recent low-level star formation in
    lower-mass model galaxies.  Time in these panels is age or
    equivalently look-back time.  `Downsizing' is clear in these
    simulated galaxies: the age of the peak of the star-formation
    history decreases with decreasing mass, although this decrease
    flattens out at the smallest masses.  A mass--metallicity relation
    is also apparent in the average metallicity distributions, with a
    `kink' and flattening in the peak metallicities around
    $M_*\sim10^{11}\,\msun$.\label{fig:tzhists}}
\end{figure}

We begin by displaying the average SFHs and metallicity distributions
of early-type galaxies in Coma cluster-sized halos in
Figure~\ref{fig:tzhists}.  In the upper panel we find clear evidence
for `downsizing' in the peak star-formation epoch of early-type
galaxies drawn from the semi-analytic models: the more massive the
galaxy, the earlier (on average) the peak in the SFH (compare with the
top panel of Fig.~1 in \citealt{deLucia06}, who found similar results
using a similar hierarchical galaxy formation model including AGN
feedback, although many of the details are different). The inset in
the upper panel shows the SFHs of the galaxies over the last 2 Gyr,
the time-scale to which \hbeta\ and the other Balmer lines are
maximally sensitive \citep[e.g.,][and Fig.~\ref{fig:bs} below]{ST07};
note that the lower the mass, the more likely the galaxy is to still
have some ongoing star formation at these late times.  In the current
models however, downsizing is not very apparent at masses below
$M_*\la10^{11}\,\msun$, and galaxies with masses
$M_*\sim10^{11}\,\msun$ have the highest fraction of recent ($t<2$
Gyr) star formation.  In the lower panel we see clear evidence for a
mass--metallicity relation in the early-type galaxies: low-mass
galaxies have lower metallicities than high-mass galaxies, as expected
\citep{Faber73}.  Furthermore it is clear that there is a `kink' in
the relation at a mass of $M_*\sim10^{11}\,\msun$ and a flattening
above this mass \citep[note the spacing of the peak of the metallicity
distributions; compare also Fig.~\ref{fig:zmass} below with Fig. 8
in][]{Gallazzi05}.

\subsection{Line strengths and SSP analysis}
\label{sec:linestrengths}

\begin{figure*}
  \includegraphics[width=132mm]{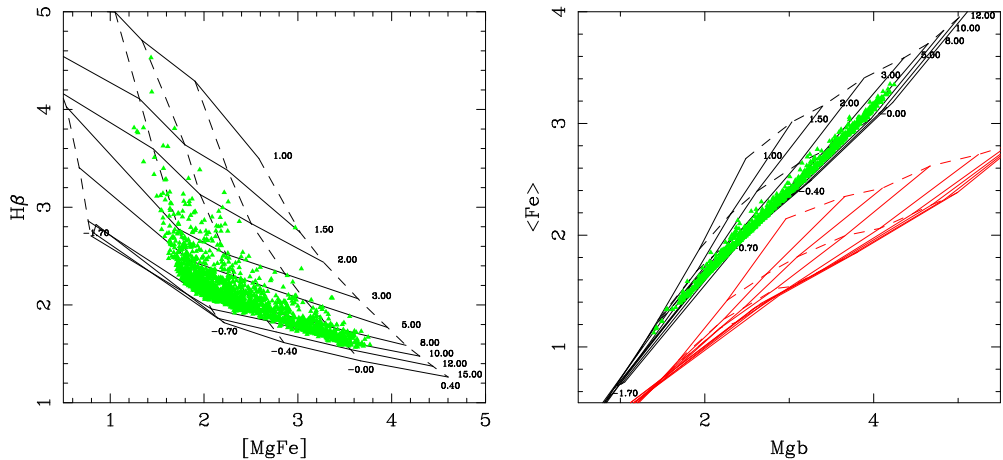}
  \caption{Line-strength indices of simulated early-type galaxies from
    20 realizations of a Coma cluster-sized halo.  Left: The
    \hbeta-\mgfe\ diagram is a convenient plot from which to read
    SSP-equivalent age (solid, roughly horizontal lines) and
    SSP-equivalent metallicity (dashed, roughly vertical lines), as
    these indices are at best only mildly affected by $\enh\neq0$
    \citep{T00a,TMK04}.  Right: The \mgb-\fe\ diagram is a convenient
    plot from which to read \enh.  The left-hand grid has $\enh=0$
    (i.e., solar composition) and the right-hand grid has $\enh=+0.3$.
    Note that these model galaxies show no evidence for $\enh\neq0$,
    as expected from the chemical evolution prescription used in the
    simulation (see text).
    \label{fig:hbmgbfe}}
\end{figure*}

Using the procedure detailed in Section~\ref{sec:stellarpops} above,
we have computed line-strength indices for simulated early-type
galaxies in Coma cluster-sized halos.  The results for \hbeta, \mgb,
and \fe\ are shown in Fig.~\ref{fig:hbmgbfe}.  The left-hand panel
shows that there is a weak trend in metallicity as a function of age,
although with a large scatter. This is in the sense that
lower-metallicity galaxies have younger ages, and originates mainly
from the trends between galaxy mass and metallicity and galaxy mass
and age in the models (low-mass galaxies are both younger and more
metal-poor). The right-hand panel of Fig.~\ref{fig:hbmgbfe} shows only
that the enhancement ratios of the model galaxies are solar, i.e.,
that $\enh=0$. This is not surprising, since we have used solar
abundance SSP models to generate the synthetic spectra.

\begin{figure*}
  \includegraphics[width=132mm]{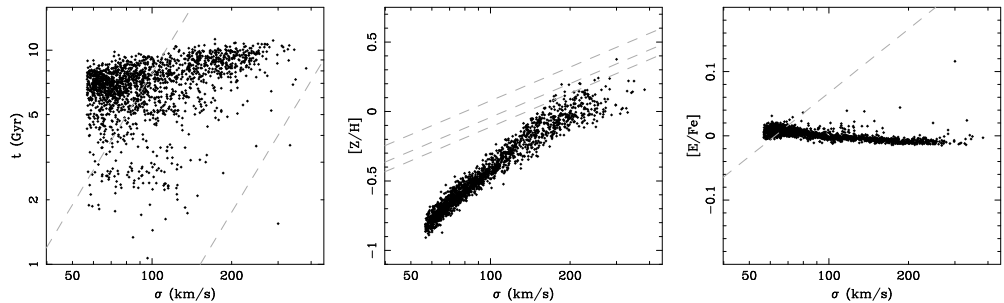}
  \caption{Velocity dispersion--stellar population parameter relations
    for model early-type galaxies selected from an ensemble of 20
    realizations of a Coma cluster-sized halo.  The dashed lines are
    projections of the observed Coma cluster $Z$ plane (for
    metallicities of $\z=-0.5$, 0, $+0.5$ dex in the left panel and
    ages of 5, 10, 15 Gyr in the middle panel) and \enh--$\sigma$
    relation (right panel) on to the relations (see \citealt{T00b};
    TFD08).  One can see `archaeological downsizing' in the left
    panel, along with a steeper mass--metallicity relation than seen
    in the data (middle panel). Note that the point with high \enh\ in
    the right-hand panel is also the object with very young age and
    high velocity dispersion ($\sigma\approx300\,\kms$) in the
    left-hand panel: it falls in a sligtly unusual region of
    line-strength space because of a burst of recent star formation
    and has a biased \enh\ as a result, even though intrinsically it
    has $\enh=0$.\label{fig:allsigmarels}}
\end{figure*}

Figure~\ref{fig:allsigmarels} shows the three derived stellar
population parameters (age, metallicity, and $\enh$) plotted against
the galaxy velocity dispersion.  The SSP-equivalent ages of the model
galaxies decrease with decreasing velocity dispersion, a phenomenon
often referred to as `archaeological downsizing' \citep[cf.][among
  others]{Nelan05,TMBO05}.  The scatter also grows with decreasing
velocity dispersion, suggesting that the formation of low-mass
galaxies is typically more extended than that of higher-mass galaxies
(as we saw in Figure~\ref{fig:tzhists}). The SSP-equivalent
metallicities show a tight and strong mass-metallicity relation, as
expected (see above), but the slope is steeper and the zeropoint is
lower than the real Coma cluster data shown by the dashed lines; we
discuss this point further below and in Sec.~\ref{sec:tzmock}.

\subsubsection{`Archaeological downsizing' and recent star formation}

\begin{figure}
  \includegraphics[width=85mm]{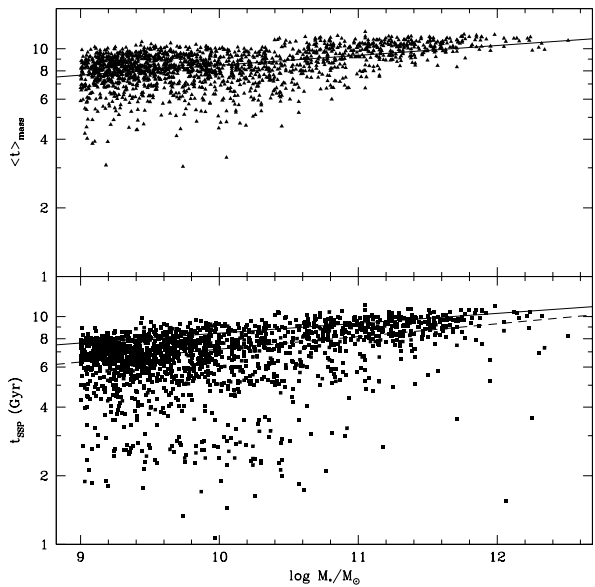}
  \caption{Mass--age relations for model early-type galaxies drawn
    from 20 realizations of a Coma cluster-sized halo.  Top:
    mass-weighted stellar age as a function of stellar mass.  The
    solid line is a $3\sigma$-clipped least-squares fit.  The relation
    is quite shallow and possibly even flat below
    $M_*=10^{10}\,\msun$.  Bottom: SSP-equivalent stellar population
    age as a function of stellar mass.  The dashed line is a
    $3\sigma$-clipped fit to these points, and the solid line is the
    fit from panel (a).  Note the large scatter in $\tssp$, the lower
    apparent ages, and the strong apparent `archaeological
    downsizing', which exaggerates the trend actually present in the
    mass-weighted ages.\label{fig:agemass}}
\end{figure}

\begin{figure}
  \includegraphics[width=85mm]{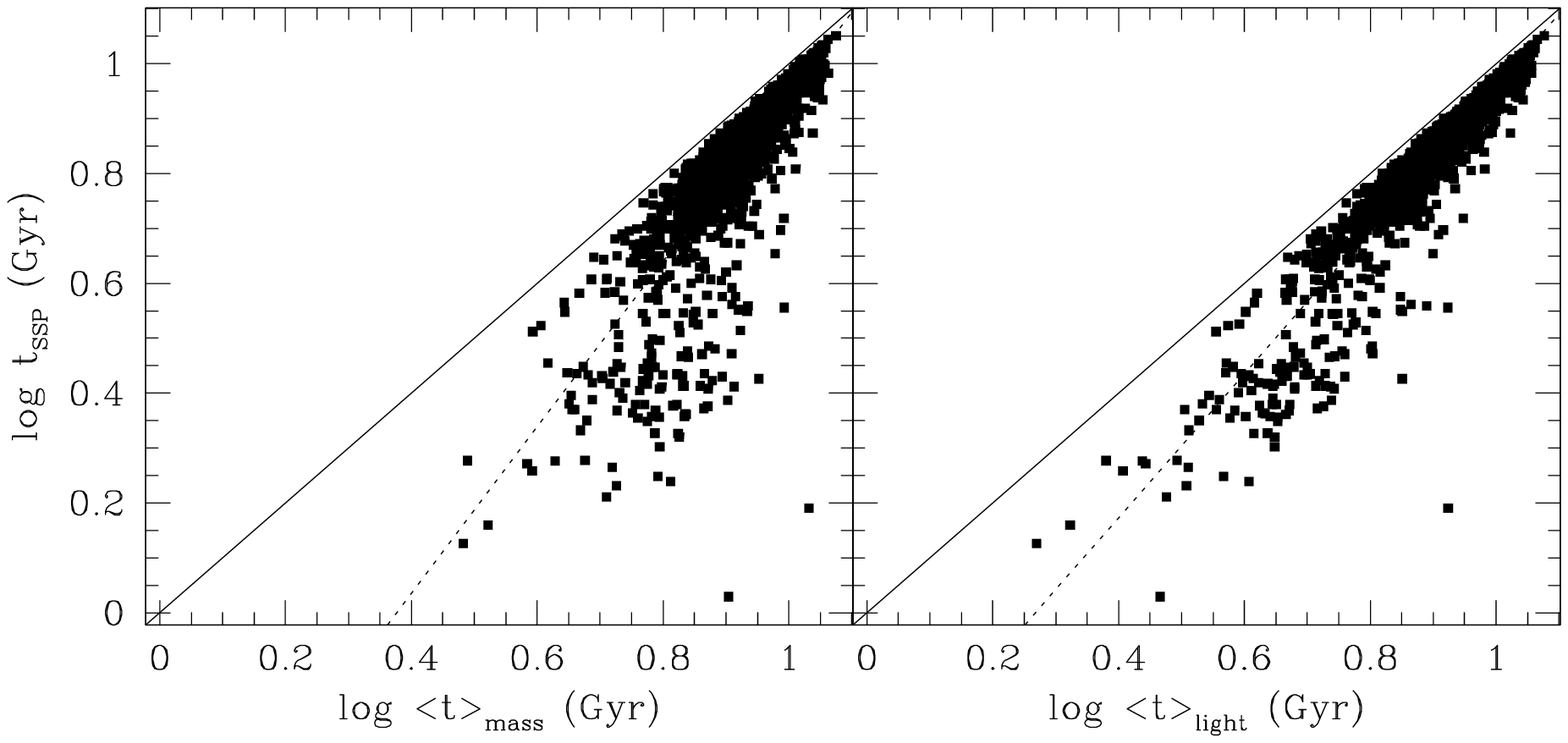}
  \caption{SSP-equivalent age as a function of mass-weighted
    (left-hand panel) and ($V$-band) light-weighted (right-hand panel)
    age for model early-type galaxies drawn from 20 realizations of a
    Coma cluster-sized halo.  Solid lines are lines of equivalence for
    the plotted parameters and dashed lines are linear-least-sequares
    fits given in Eqs.~\ref{eq:tmass} (left-hand panel) and
    \ref{eq:tlight} (right-hand panel).  The SSP-equivalent and
    light-weighted ages are correlated, as are the SSP-equivalent and
    mass-weighted ages, but the scatter is very large, the slope is
    not unity, and the SSP-ages are biased low. The scatter is larger
    for the more physically interesting mass-weighted ages. These
    biases are due to residual star formation in some fraction of the
    early-type galaxies \citep[cf.][]{ST07}.\label{fig:tcomp}}
\end{figure}

Given the known sensitivity of Balmer lines to recent star formation
\citep[recent star formation: RSF, e.g.,][]{T00b,ST07}, we ask whether
the `archaeological downsizing' seen in these models is a reflection
of true downsizing in the mass-weighted ages of the (model) galaxies
or just results from small amounts of star formation occurring
preferentially in smaller galaxies with extended star-formation
histories.  In Figure~\ref{fig:agemass}, we see that the latter case
is likely to be the dominant signal in `archaeological downsizing' in
these models: low-mass galaxies have more recent star formation than
high-mass galaxies, as expected from Figure~\ref{fig:tzhists},
resulting in younger SSP-equivalent ages.  This can be seen even more
clearly in the left-hand panel of Figure~\ref{fig:tcomp}, which shows
that the SSP-equivalent ages of galaxies are \emph{always} younger
than the mass-weighted ages, with the discrepancy (and scatter)
growing towards younger -- and thus lower-mass, following
Figure~\ref{fig:agemass} -- galaxies.  We therefore suggest that
`archaeological downsizing' will always appear stronger than any true
(mass-weighted) downsizing in the stellar populations due to the bias
caused by recent star formation in low-mass galaxies.  This is not to
say that downsizing does not happen in these models -- it clearly
does, as shown in the top panel of Figure~\ref{fig:agemass} -- rather
that `archaeological downsizing' results \citep[such
  as][]{Nelan05,TMBO05} almost certainly overstate the true magnitude
of downsizing, as suggested by TFD08.

Further examining Figure~\ref{fig:tcomp}, we find that the
SSP-equivalent ages of model galaxies are in fact correlated with both
the mass- and light-weighted ages of the galaxies, but the scatter is
large in both cases (especially for the mass-weighted ages), as
expected from the preceeding discussion.  Simple linear fits to the
points in Figure~\ref{fig:tcomp} give the following relations:
\begin{eqnarray}
  \log\tssp &=& 1.51\,(\pm0.02)\,\log\langle t\rangle_{\mathrm{mass}} -
  0.568\,(\pm0.002) \label{eq:tmass}\\
  \log\tssp &=& 1.31\,(\pm0.01)\,\log\langle t\rangle_{\mathrm{light}}
  - 0.351\,(\pm0.001) \label{eq:tlight}
\end{eqnarray}
for 1801 model galaxies, where the uncertainties are formal $1\sigma$
errors assuming equal weighting of points in the least-squares fits
and the RMS scatters are 0.08 and 0.05 dex, respectively.  We note the
following points.
\begin{itemize}
\item As discussed above, RSF strongly biases the inferred ages, such
that SSP-equivalent ages are younger than mass-weighted ages by more
than 40 per cent on average and younger than light-weighted ages by
roughly 25 per cent on average, thus greatly exaggerating the effect
of downsizing on the stellar population ages.
\item The scatter is asymmetric in both diagrams.  As discussed by
\citet{ST07}, galaxies that appear young in \tssp\ are not guaranteed
to be -- and in these simulations, unlikely to be -- young in a mass-
or light-weighted sense.
\item The biases can be expected to be even worse for smaller samples
  of galaxies. As an example, we consider the relations between
  SSP-equivalent and mass- or light-weighted age for a mock catalogue
  representing the \citet{M02} sample of Coma cluster galaxies (see
  Sec.~\ref{sec:mockcats} below).  For this sample of 108 (model)
  galaxies, the slopes of the relations are steeper [$\log\tssp
    \propto (1.89\pm0.25) \log\langle t\rangle_{\mathrm{mass}}$ and
    $\log\tssp \propto (1.53\pm0.15) \log\langle
    t\rangle_{\mathrm{light}}$] and the scatters are larger (0.18 and
  0.16 dex, respectively).
\end{itemize}

\begin{figure}
  \includegraphics[width=85mm]{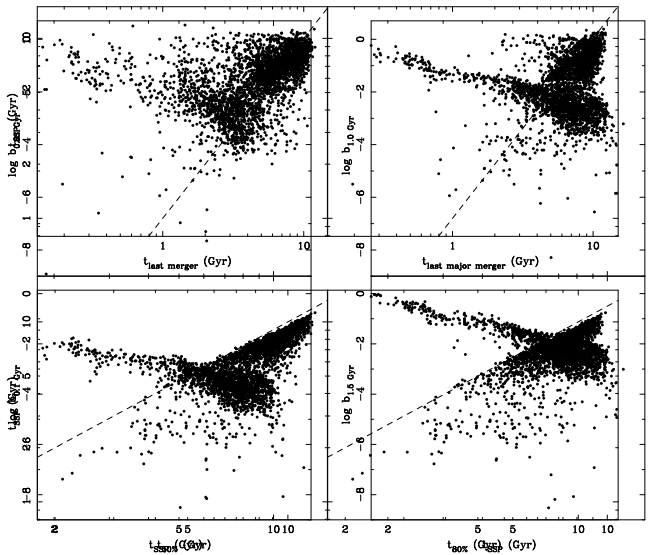}
  \caption{Comparison of SSP-equivalent age with galaxy assembly time
    scales.  Top panels: \tssp\ as a function of the time of last
    merger (left-hand panel) and the time of last major merger
    (right-hand panel).  Bottom panels: \tssp\ as a function of the
    time at which 50 per cent of stellar mass was formed (left-hand
    panel) and the time at which 80 per cent of stellar mass was
    formed (right-hand panel).\label{fig:mergers}}
\end{figure}

If the SSP-equivalent age of a galaxy does not accurately represent
its mass- or light-weighted age, is there some other relevant
time-scale of its star-formation history that it does reflect?  Using
our knowledge of the mass- and star-formation histories of the model
galaxies, in Figure~\ref{fig:mergers} we present \tssp\ as a function
of the times of the last merger (mass ratio greater than 1:10)
and the last major merger (mass ratio greater than 1:5), and the times
at which 50 per cent and 80 per cent of the present-day stars were
formed.  The correlations of \tssp\ with these quantities is even
worse than with the mass- or light-weighted ages.  In the case of the
merger times, recent dry mergers leave \tssp\ relatively unchanged,
while minor but gas-rich mergers can seriously perturb \tssp.  The
lack of a strong correlation of \tssp\ with the times at which 50 and
80 per cent of the present-day stars formed is due to again to the
presence of RSF.  If the last small fraction of stellar mass is formed
recently in a galaxy, this RSF results in a young \tssp.

\begin{figure}
\includegraphics[width=85mm]{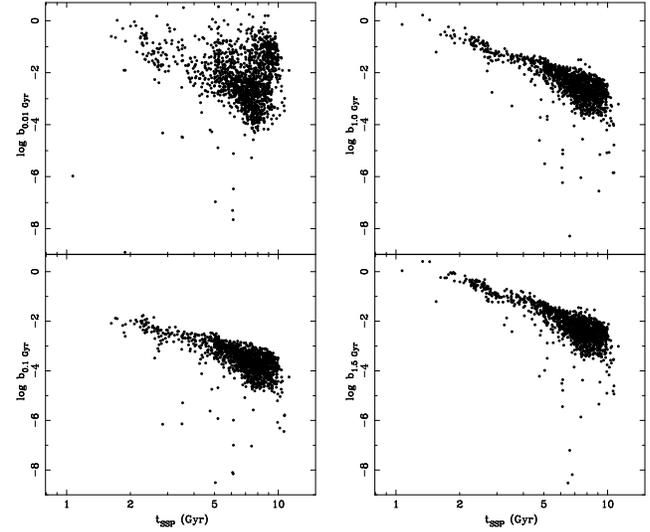}
\caption{Correlation of SSP-equivalent age and `birthrate' parameter,
  which is the star formation rate averaged over a relatively short
  timescale $\tau$, divided by the star formation rate over the whole
  lifetime of the galaxy. Top left-hand panel: birthrate with
  $\tau=10$ Myr ($b_{0.01\,\mathrm{Gyr}}$).  Bottom left-hand panel:
  birthrate with $\tau=100$ Myr ($b_{0.1\,\mathrm{Gyr}}$).  Top
  right-hand panel: birthrate with $\tau=1.0$ Gyr
  ($b_{1.0\,\mathrm{Gyr}}$).  Bottom right-hand panel: birthrate with
  $\tau=1.0$ Gyr ($b_{1.5\,\mathrm{Gyr}}$).  (Galaxies with no star
  formation within the appropriate time-scale are not plotted.)  The
  correlation of \tssp with $b_{0.1\,\mathrm{Gyr}}$,
  $b_{1.0\,\mathrm{Gyr}}$, and $b_{1.5\,\mathrm{Gyr}}$ highlights the
  sensitivity of the SSP-equivalent age to recent star
  formation.\label{fig:bs}}
\end{figure}

Rather than a specific event in the history of the galaxy, the
SSP-equivalent age represents a rough measure of the birth-rate of
stars over the last 1--1.5 Gyr.  This can be seen in
Figure~\ref{fig:bs}, where we plot the average birth-rate $b_t$ -- the
fraction of stellar mass formed over some time with respect to the
total stellar mass -- for four time-scales: 0.01, 0.1, 1.0, and 1.5
Gyr.  We see that the best correlation of \tssp\ with $b_t$ is when
$1<t<1.5$ Gyr, suggesting that SSP-equivalent ages inferred from
\hbeta\ are maximally sensitive to RSF in the past Gyr \citep[as shown
from simpler, two-burst models in][]{ST07}.  We therefore conclude
that \tssp\ is mostly measuring RSF over $\sim$ 1-Gyr timescales, not
mass- or light-weighted ages, not merger times and not the time at
which some (large) fixed fraction of stellar mass has formed.

\subsubsection{Metallicities}

\begin{figure}
\includegraphics[width=85mm]{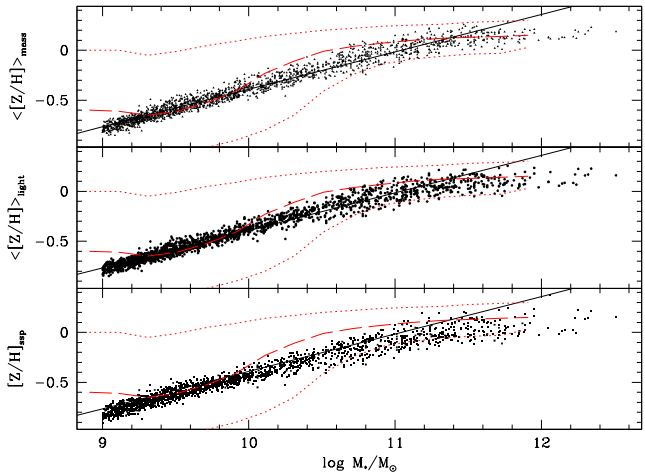}
\caption{Mass--metallicity relations for model early-type galaxies
  drawn from 20 realizations of a Coma cluster-sized halo.  Top:
  mass-weighted metallicity as a function of stellar mass.  The solid
  line is a $3\sigma$-clipped least-squares fit to the relation.
  There is clearly a very strong, tight mass--metallicity relation
  which flattens out at masses above $M_*\sim3\times10^{11}\,\msun$.
  The long-dashed (red) line is the mass--$r$-band-light-weighted
  metallicity relation from the SDSS sample of \citet{Gallazzi05} and
  the dotted (red) lines are the $1\sigma$ marginalised confidence
  bounds on this relation.  Middle: $V$-band light-weighted
  metallicity as a function of stellar mass.  Solid, long-dashed and
  dotted lines are the same as in the top panel.  Note that the model
  light-weighted metallicities are slightly too low compared with the
  SDSS galaxies.  Bottom: SSP-equivalent stellar population
  metallicity as a function of stellar mass.  Solid, long-dashed and
  dotted lines are again the same as in the top panel.  The scatter is
  slightly larger, the slope is slightly flatter and the zero point is
  slightly lower than for the mass--mass-weighted metallicity
  relation.\label{fig:zmass}}
\end{figure}

We turn briefly to the metallicities of the model galaxies.  We
present their mass--metallicity relations in Figure~\ref{fig:zmass}
both for the mass-weighted and SSP-equivalent metallicities.  As
expected from the bottom panel of Figure~\ref{fig:tzhists}, a distinct
flattening of the relations occurs at masses above
$\log(M_*/\msun)\ga11.3$.  This is in rough accordance with the
results of \citet{Gallazzi05}, who found a similar effect in the
$r$-band-luminosity-weighted metallicities of SDSS galaxies (see also
Fig. 6 of S08).  Our ($V$-band) light-weighted metallicities are
slightly too low compared with the data; the difference in the scatter
is due to observational uncertainties in the SDSS data that we have
not attempted to model.  However, our intent here is not to make a
detailed comparison with the SDSS data; rather, we will generate mock
catalogues to compare with the Coma cluster galaxies in
Section~\ref{sec:tzmock} below.

\begin{figure}
\includegraphics[width=85mm]{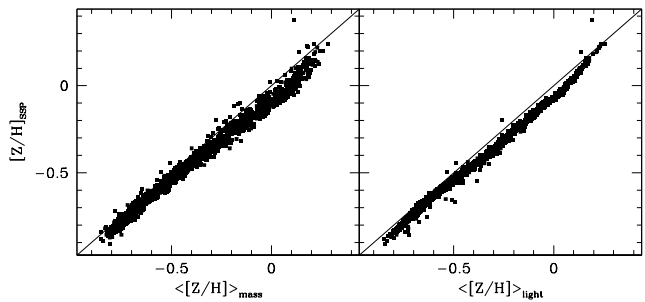}
\caption{SSP-equivalent metallicity as a function of mass-weighted
  (left-hand panel) and ($V$-band) light-weighted (right-hand panel)
  metallicity for model early-type galaxies drawn from 20 realizations
  of a Coma cluster-sized halo.  Solid lines are lines of equivalence
  for the plotted parameters.  The SSP-equivalent metallicities
  correlate well with mass- and light-weighted metallicities, albeit
  with a small offset.\label{fig:zcomp}}
\end{figure}

We compare mass- and light-weighted metallicities with SSP-equivalent
metallicities in Figure~\ref{fig:zcomp}.  Unlike in the case of \tssp,
\zssp\ correlates almost perfectly with mass- and light-weighted
metallicity \citep[as shown in][]{ST07}, with very small scatter in
the latter case.  The SSP-equivalent metallicity is therefore a very
good tracer of the light- or even mass-metallicity of a galaxy, modulo
a small offset towards lower metallicity.

\begin{figure}
  \includegraphics[width=85mm]{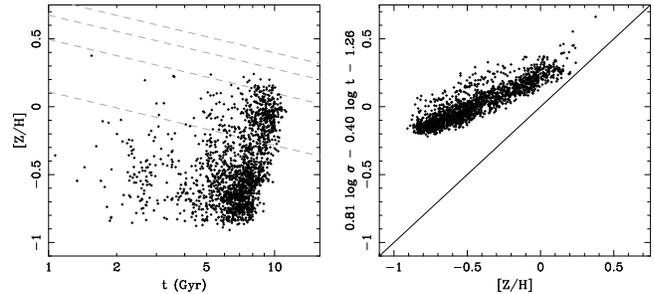}
  \caption{A nearly face-on (left-hand panel) and the edge-on
    (right-hand panel) view of the $Z$ plane \citep{T00b} for model
    galaxies from the ensemble of 20 realizations of a Coma
    cluster-sized halo.  The dashed lines in the left panel are
    projections of the observed Coma cluster $Z$ plane (TFD08) onto
    the age--metallicity plane for $\sigma=50$, $150$, $250$,
    $350\,\kms$.  The solid line in the right panel is the edge-on
    view of the $Z$ plane.  At a given velocity dispersion, model
    metallicities are typically too low with too steep of a slope, as
    seen in the middle panel of Fig.~\ref{fig:allsigmarels}.
  \label{fig:allzplane}}
\end{figure}

Finally, we turn to the $Z$ plane of \citet{T00b}, a relation between
velocity dispersion and SSP-equivalent age and metallicity.  In
Figure~\ref{fig:allzplane} we present these quantities projected onto
the $Z$ plane of Coma cluster early-type galaxies determined in TFD08.
The left-hand panel shows that the metallicities of the model galaxies
are too low at a given velocity dispersion, while the right-hand panel
shows that the slope of the edge-on view of the $Z$ plane of the model
galaxies is too shallow compared with the observations.  The former
problem is due to the chemical evolution model used and assumptions
about the IMF.  Arrigoni et al.\ (in prep.) have shown that a more
detailed, multi-element chemical evolution model that includes
enrichment by Type Ia supernovae and a slight flattening of the IMF
(to produce more high-mass stars and therefore more metals) can remove
this discrepancy.  The latter problem is either due to the wrong
mass--metallicity slope (which seems unlikely from
Fig.~\ref{fig:zmass}) or to a different correlation between \tssp\ and
\zssp\ in the data and models.  We consider these possibilities in
more detail below.

\section{Coma cluster: observations and mock catalogues}
\label{sec:coma}

We now to turn to our second and third goals, which are interpreting
the properties of local early-type cluster galaxies in the context of
hierarchical formation models and testing the models themselves using
the data.

\subsection{Colour--magnitude diagram}
\label{sec:cmd}

\begin{figure}
  \includegraphics[width=85mm]{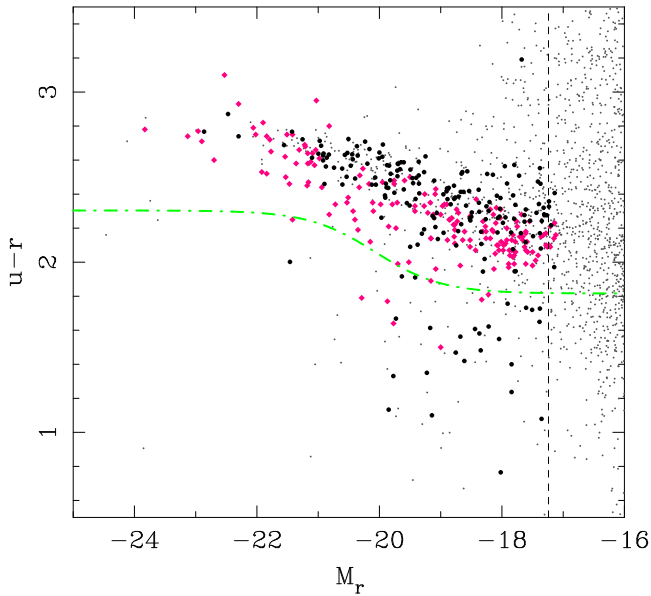}
  \caption{Observed and predicted colour--magnitude diagram (CMD) of
    the Coma cluster.  Large black dots show galaxies with redshifts
    from the Sloan Digital Sky Survey (SDSS) DR6 \citep{DR6}, placing
    them at the distance of the Coma cluster; small grey points are
    other galaxies in the same field without DR6 redshifts, assuming
    that they are also in the cluster; and (pink) diamonds are model
    galaxies drawn from a single realization of a Coma cluster-sized
    halo, with no regard to morphology or colour.  The dot-dashed
    green line demarcates the blue-red galaxy division of
    \citet{Baldry04}; note the excellent agreement of the model and
    observed red sequence slope and good agreement of the zero-points.
    The dashed vertical line represents the magnitude limit of SDSS
    spectroscopy, which is close to the model-galaxy mass selection
    limit of $10^9\,M_{\odot}$.\label{fig:cmd}}
\end{figure}

To begin our confrontation between the models and data, we ask whether
the current galaxy formation models can reproduce the
colour--magnitude diagram (CMD) of galaxies -- regardless of
morphology or colour -- in the Coma cluster.  In Figure~\ref{fig:cmd}
we show the magnitudes and colours of a galaxies drawn from a single
Coma-sized halo overlaid on the CMD of Coma cluster galaxies drawn
from the sixth data release of the Sloan Digital Sky Survey \citep[DR6
  of SDSS][]{DR6}.  We divide the CMD into red and blue regions using
the magnitude-dependent cut of \citet{Baldry04}, which provides a
fairly clean separation of both the model and Coma cluster galaxies.
We see that the slope of the red-sequence of model galaxies matches
the Coma cluster galaxies very well.  The model galaxies appear to be
slightly too blue, which is likely due to the too-low metallicities of
galaxies in the model.  There are also not enough truly blue galaxies
down to the limits of the model and the spectral data.  However, this
may be due to the fact that the model galaxies are selected down to a
fixed stellar mass, not a given luminosity.  As blue galaxies have low
mass-to-light ratios, star-forming galaxies with masses below our mass
limit can have luminosities higher than red galaxies above our mass
limit.  The good (if not perfect) match between the model and observed
CMDs give us confidence to proceed with attempting to match detailed
stellar population properties. We note that the luminosity function
predicted by the model is also in good agreement with the observed
Coma luminosity function.

\subsection{Observational material}
\label{sec:data}

\begin{table*}
  \begin{minipage}{132mm}
  \caption{Coma cluster observations}
  \label{tbl:comaobs}
  \begin{tabular}{lrrrrr}
    \hline \multicolumn{1}{c}{Sample}&
    \multicolumn{1}{c}{$N_{\mathrm{gal}}$}&
    \multicolumn{1}{c}{$\sigma_{\hbeta}^{200}$ (\AA)}&
    \multicolumn{1}{c}{$\sigma_{\mgb}^{200}$ (\AA)}&
    \multicolumn{1}{c}{$\sigma_{\mathrm{Fe5270}}^{200}$ (\AA)}&
    \multicolumn{1}{c}{$\sigma_{\mathrm{Fe5335}}^{200}$ (\AA)}\\ 
    \hline
    TFD08&12&0.032&0.033&0.037&0.041\\
    \citet{M02}&121&0.096&0.095&0.102&0.163\\
    \citet{Nelan05}&100&0.074&0.080&0.091&0.105\\
    \hline
  \end{tabular}

  Cols. 3--6: Mean index uncertainties for a galaxy with
  $\sigma=200\,\kms$.
  \end{minipage}
\end{table*}

To compare the hierarchical galaxy formation models with observed
stellar population parameters of early-type Coma cluster galaxies, we
use the data presented in TFD08, concentrating on three samples: the
LRIS data of TFD08 (hereafter `the LRIS sample'), which has excellent
quality (in terms of signal-to-noise) but only twelve galaxies; the
large, morphologically-selected, complete sample of \citet[hereafter
  `the Moore sample']{M02}, of somewhat poorer quality than the LRIS
sample but containing ten times as many galaxies; and the large,
red-sequence-selected sample of \citet[hereafter `the Nelan
  sample']{Nelan05}, again of poorer quality than the LRIS sample but
again much larger.  Characteristics of these three samples are
presented in Table~\ref{tbl:comaobs}.  

The line strengths of the LRIS sample galaxies have been adjusted to
represent those taken through an aperture with radius equal to
one-fourth of the effective radius ($r=r_e/4$) of each galaxy
\citep[see][for more details on the method, and note that the
  apertures in that paper are denoted in diameters, not radii]{TFD08}.
This has been done in an attempt to fairly compare the `global'
stellar populations of the Coma cluster galaxies with the simulated
galaxies, which are unresolved and therefore should be considered to
represent `global' properties.  The $r_e/4$ aperture does not
perfectly represent the `global' properties, but we show in the
Appendix that stellar populations observed through apertures with
radii $r=r_e$ are typically negligibly older and $0.10$ dex more
metal-poor.  The line strengths of the Moore and Nelan samples were
taken through fiber diameters of $2\farcs7$ and $2\farcs0$; the
typical `global' ages and metallicities (i.e., taken through an
aperture of radius $r_e$) for these apertures are again negligibly
older and $0.10$ dex more metal-poor.  We will refer to these shifts
from observed to `global' parameters below.

\subsection{Mock catalogues}
\label{sec:mockcats}

\begin{figure}
  \includegraphics[width=60mm]{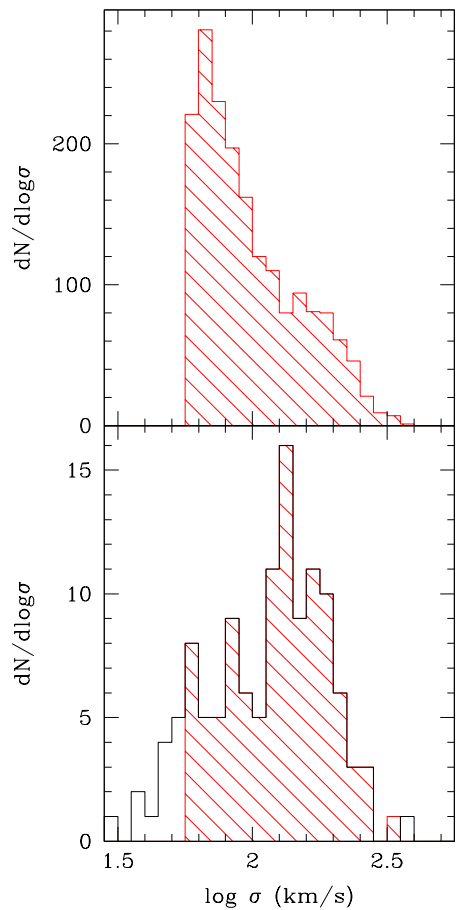}
  \caption{Top panel: Velocity-dispersion function (VDF) of model
    galaxies for a suite of twenty Coma cluster-sized haloes.  Bottom
    panel: VDF of observed Coma cluster galaxies in the \citet{M02}
    sample (black) and of the mock catalogue built to match the
    \citet{M02} galaxies (red hashed). Note that the VDF of the full
    suite is missing galaxies at the extrema of the observed VDF, and
    so these are also missing from the mock catalogue (see
    text).\label{fig:vdf}}
\end{figure}

To make the desired comparison in a sensible way, we construct mock
catalogues of early-type galaxies selected from the simulations.
These mock catalogues consist of model galaxies selected to be
early-type galaxies by the criteria discussed in \S\ref{sec:mergers}
above or red-sequence galaxies (see below) and to have the same
\emph{velocity-dispersion function} (VDF) as the observed galaxies.  A
VDF is merely the histogram of (logarithmic) velocity dispersions
$\log\sigma$ with a specified bin width; in this paper we use a bin
size of $\Delta\log\sigma=0.05$ dex.  As much as is possible, the mock
catalogue is constructed such that each bin in the VDF is sampled from
the simulated galaxies with the appropriate velocity dispersions
without replacement.  Only when a VDF bin is less populated in the
simulation than in the observations are galaxies selected from the
simulation more than once (although these galaxies will have different
index strengths due to the assumed index errors; see below).  If there
are no simulated galaxies in a VDF bin populated by the observed
galaxies, as can be the case at the highest (typically cD galaxies)
and lowest velocity dispersions, no galaxies are added to the mock
catalogue.  As noted in Section~\ref{sec:sigma}, the mass limit of our
simulations is $M_*>10^9\,\msun$, corresponding to $\sigma>57\,\kms$.
Galaxies with velocity dispersions lower than this limit are not
included in the mock catalogues.  An example of the result of our
selection algorithm is given in Figure~\ref{fig:vdf} for the mock
catalogue of the \citet{M02} sample.

The Nelan sample is a colour- and magnitude-limited sample designed to
select red-sequence galaxies \citep{Smith04}.  Galaxies in this sample
were selected to have $R_c<17$ mag and colours redder than
$\Delta(B-R_c)>-0.2$, where $\Delta(B-R_c)$ is the offset in colour at
fixed magnitude from the colour--magnitude relation of the red
sequence.  The red-sequence colour--magnitude zero-point was a median
fit to all galaxies in a given cluster with $R_c<16$ mag, assuming a
red-sequence slope of $-0.06\,\mathrm{mag}\,\mathrm{mag^{-1}}$.  These
selection criteria were also applied to the model galaxies, ignoring
the brighter limit of the zero-point fitting, which in practice makes
little difference.  One complication is that, for simplicity, $B$ and
$R_c$ magnitudes were not computed directly in the models, but were
instead determined from SDSS $gri$ magnitudes using the
transformations given in \citet{BR07}.  Again, this makes little
difference in practice.  Note that these red-sequence galaxies can
have any morphology and are not selected to be early-type galaxies.

For each galaxy selected from the simulation to populate the VDF, the
average errors in the line-strength indices of the observed galaxies
\emph{in that bin} are applied to the indices of that galaxy to
determine its `observed' indices and are labelled as its line-strength
index errors.  SSP-equivalent stellar population parameters are then
determined from the `observed' indices as described in
Sec.~\ref{sec:stellarpops}.

\subsection{Mock catalogues: comparison with observations}
\label{sec:tzmock}

\begin{figure}
  \includegraphics[width=85mm]{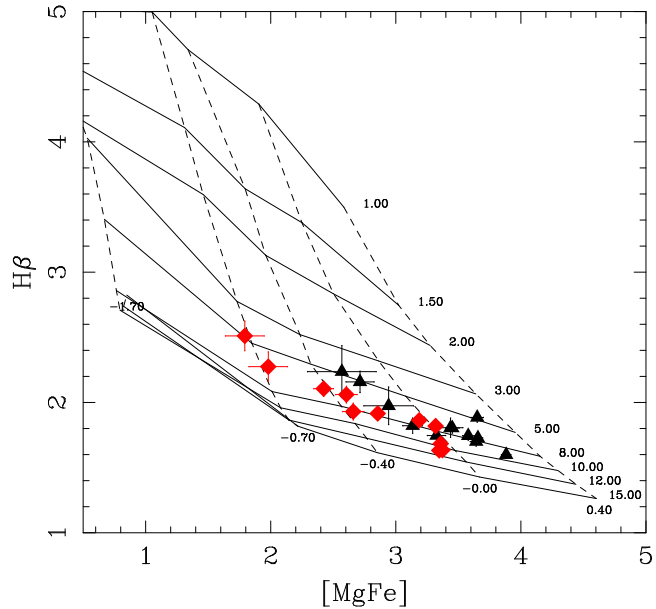}
  \caption{Observed (black) and mock (red) line strengths of the TFD08
    sample.  Model grids are the same as in Fig.~\ref{fig:hbmgbfe}.
    Note that the mock galaxies have a similar mean age (\hbeta) as
    the bulk of observed sample -- although the observed scatter is
    larger -- but the metallicities (\mgfe) are shifted too
    low.\label{fig:scthbmgbfe}}
\end{figure}

\begin{figure}
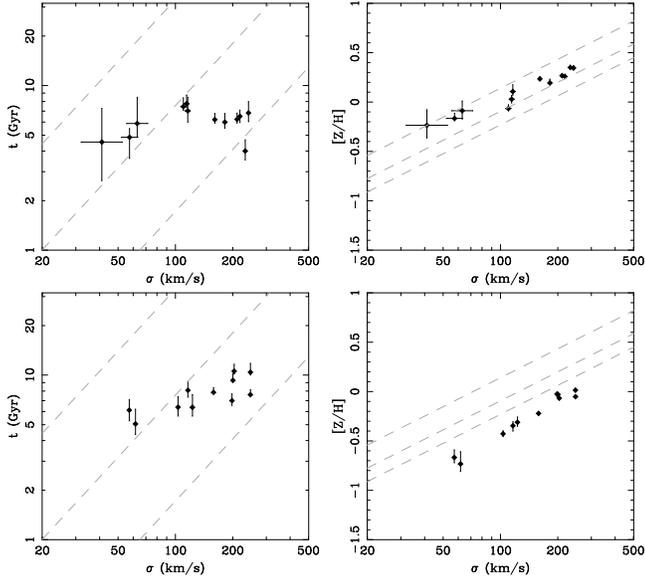

  \includegraphics[width=85mm]{agedate_ls_bc03newalpha_comaapre2gi_hbmgbfe_newerrs-sigmas.eps}

  \includegraphics[width=85mm]{tze_sct_coma_sams_bc03_sf_stpops-sigmas.eps}
  \caption{Velocity dispersion--stellar population relations for the
    observed (top panels) and mock (bottom panels) TFD08 sample (see
    Fig.~\ref{fig:allsigmarels}).  As in Fig.~\ref{fig:scthbmgbfe},
    the age distribution of the mock galaxies is similar to the
    observed galaxies (left-hand panels), although a bit steeper as a
    function of velocity dispersion, while the metallicities of the
    mock galaxies are too low (right-hand panels).  This difference in
    metallicity is \emph{not} entirely due to aperture effects, which
    can contribute only $\sim0.1$ dex of the difference (see text and
    the Appendix).  The dashed lines are the projections of the
    observed Coma cluster $Z$ plane (see Figs.~\ref{fig:allsigmarels}
    and \ref{fig:allzplane}).\label{fig:sctsigmarels}}
\end{figure}

\begin{figure}
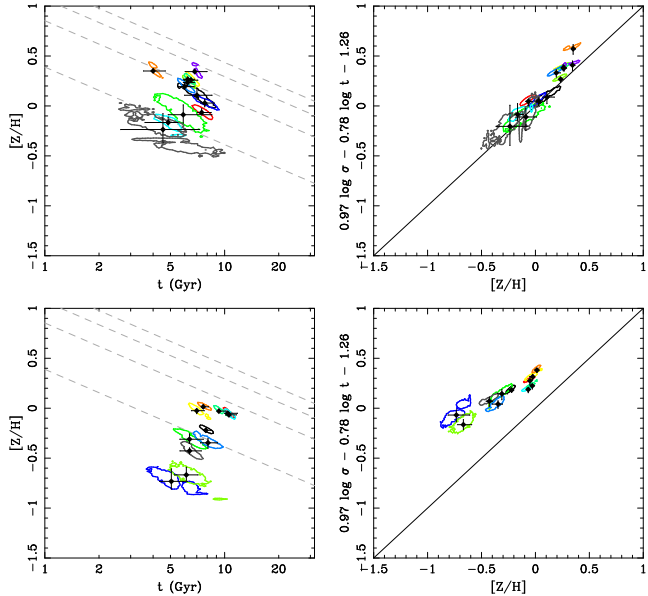

  \includegraphics[width=85mm]{agedate_ls_bc03newalpha_comaapre2gi_hbmgbfe_newerrs-zplane.eps}

  \includegraphics[width=85mm]{tze_sct_coma_sams_bc03_sf_stpops-zplane.eps}
  \caption{The nearly face-on (left-hand panels) and edge-on
    (right-hand panels) views of the $Z$ plane \citep{T00b} in the
    observed (top panels) and mock (bottom panels) TFD08 sample.
    Contours are 68 per cent confidence intervals; note that the
    age--metallicity correlations are accentuated by but do not arise
    from the age--metallicity degeneracy. As in
    Figs.~\ref{fig:scthbmgbfe} and \ref{fig:sctsigmarels}, it is clear
    that the metallicity of the mock catalogue is too low, but the age
    distribution is similar.  Again, the difference in metallicity is
    not entirely due to aperture effects (see text).  The dashed lines
    are the projections of the observed Coma cluster $Z$ plane (see
    Figs.~\ref{fig:allsigmarels} and
    \ref{fig:allzplane}).\label{fig:sctzplane}}
\end{figure}

We begin by comparing the predictions of the models with the TFD08
LRIS sample.  The predicted and observed line strengths are shown in
Figure~\ref{fig:scthbmgbfe}.  We see that the \hbeta\ strengths of the
observed and model galaxies are quite similar, but the
\mgfe\ strengths are too low.  Figure~\ref{fig:sctsigmarels} shows
that this discrepancy is driven by the metallicity offset mentioned
above -- the metallicities of the model galaxies are too low by
$\sim0.4$ dex and have a too-steep velocity-dispersion--metallicity
relation (right-hand panels).  The difference in metallicities is
\emph{not} due entirely to aperture effects, as we have shown in the
Appendix that such effects can only change the metallicity by
$\sim0.1$ dex.  The left-hand panels of Figure~\ref{fig:sctsigmarels}
show however that the observed and model (SSP-equivalent) ages are
similar, although the models show a weak `archaeological downsizing'
effect that is not clearly present in the observations.  Note that
TFD08 claim that there is \emph{no} `archaeological downsizing' in
this sample, but this is admittedly a very small sample.  Examining
the $Z$ plane in Figure~\ref{fig:sctzplane}, we find that, apart from
the obvious metallicity offset, the current galaxy formation models
produce a similar $Z$ plane to that observed.  However, the slight
correlation between \tssp\ and \zssp\ in the models seen in the bottom
left-hand panel of this figure may be the cause of the too-steep
velocity-dispersion--metallicity relation seen in
Figure~\ref{fig:sctsigmarels}. Again, the LRIS sample is very small,
but the quality of the data suggests that a large survey of a cluster
like Coma with very high-quality data would be very instructive to
probe these models even further; we pursue this idea further below.

\begin{figure}
  \includegraphics[width=85mm]{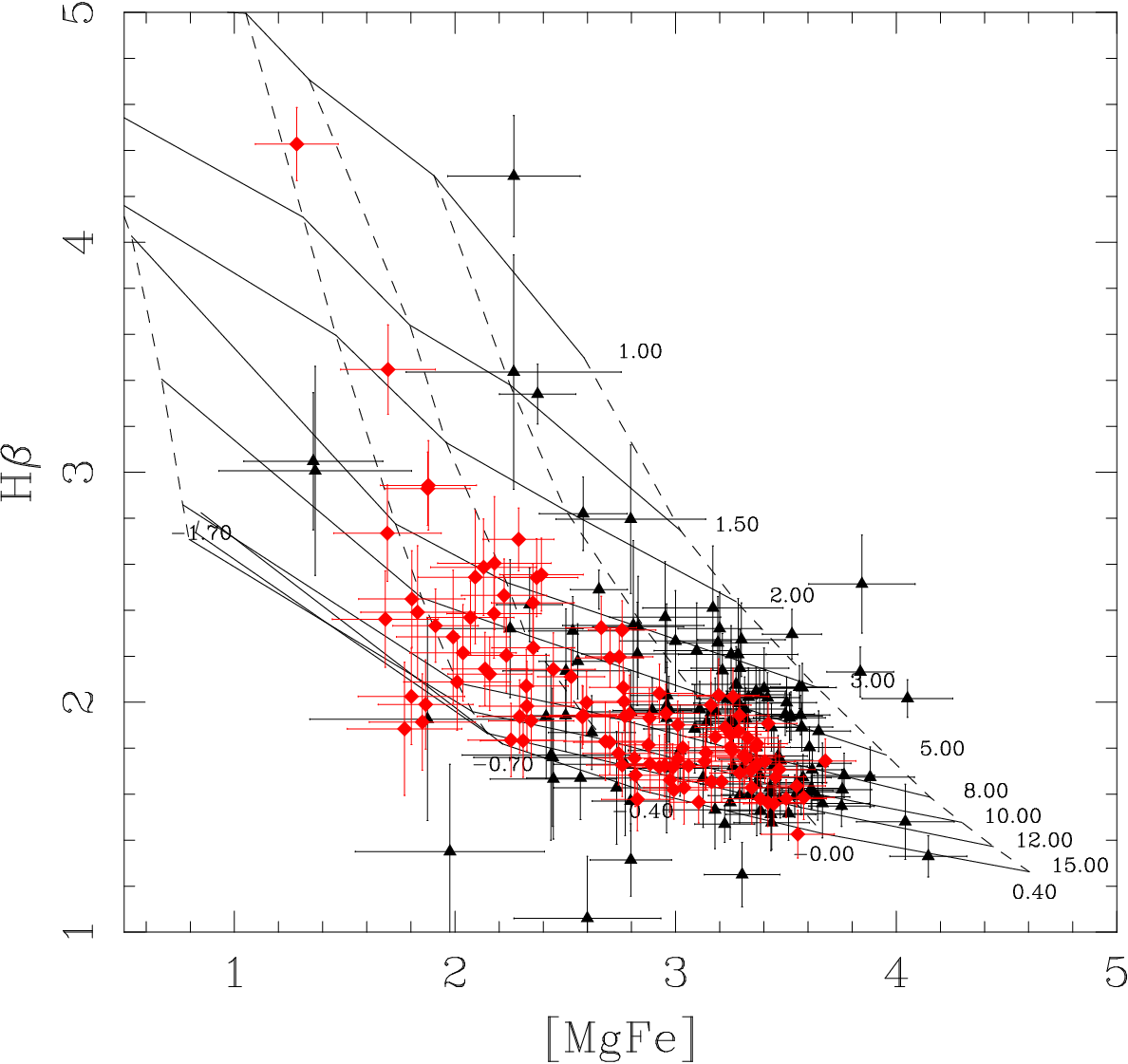}
  \caption{As for Fig.~\ref{fig:scthbmgbfe}, but for the \citet{M02}
  sample.\label{fig:moorehbmgbfe}}
\end{figure}

\begin{figure}
  \includegraphics[width=85mm]{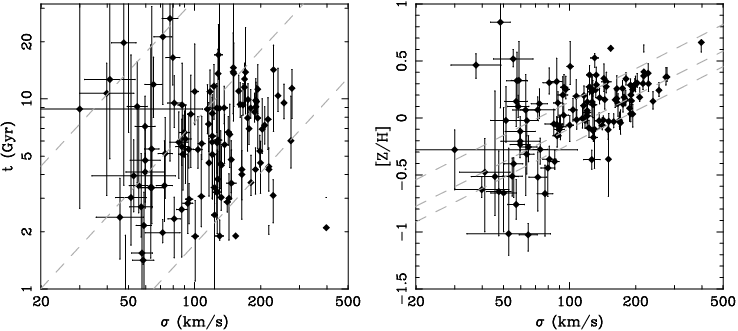}

  \includegraphics[width=85mm]{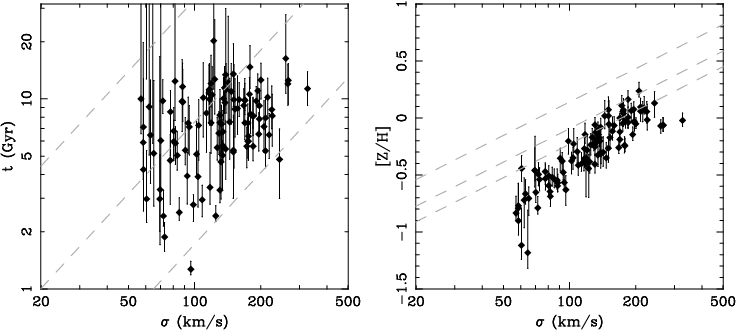}
  \caption{As for Fig.~\ref{fig:sctsigmarels}, but for the \citet{M02}
  sample.\label{fig:mooresigmarels}}
\end{figure}

\begin{figure}
  \includegraphics[width=85mm]{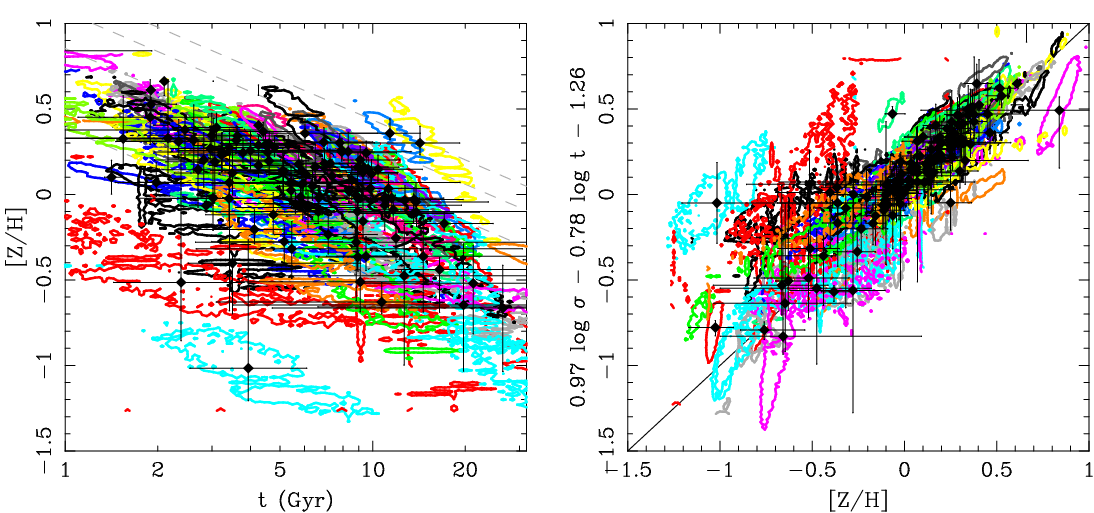}

  \includegraphics[width=85mm]{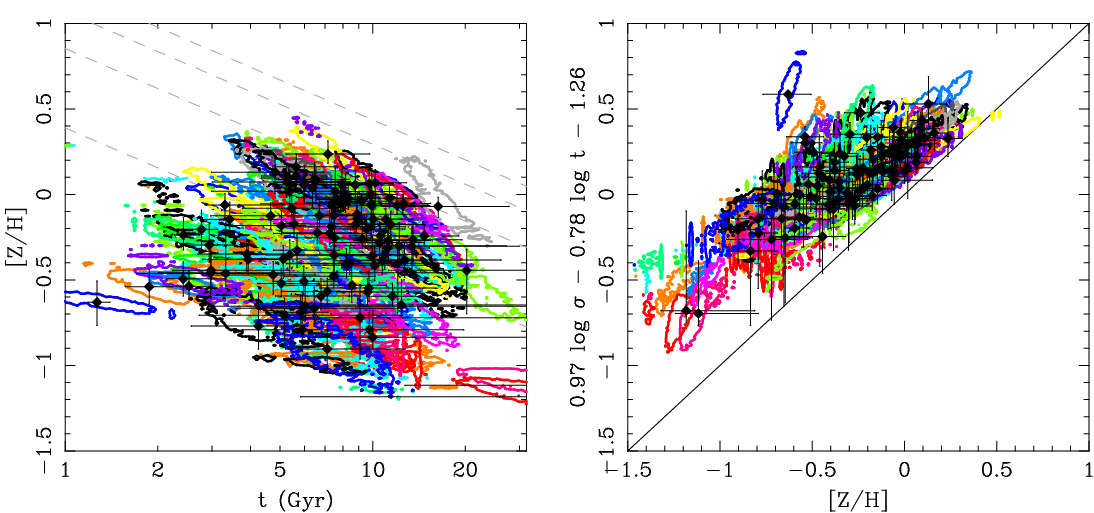}
  \caption{As for Fig.~\ref{fig:sctzplane}, but for the \citet{M02}
  sample.\label{fig:moorezplane}}
\end{figure}

\begin{figure}
  \includegraphics[width=85mm]{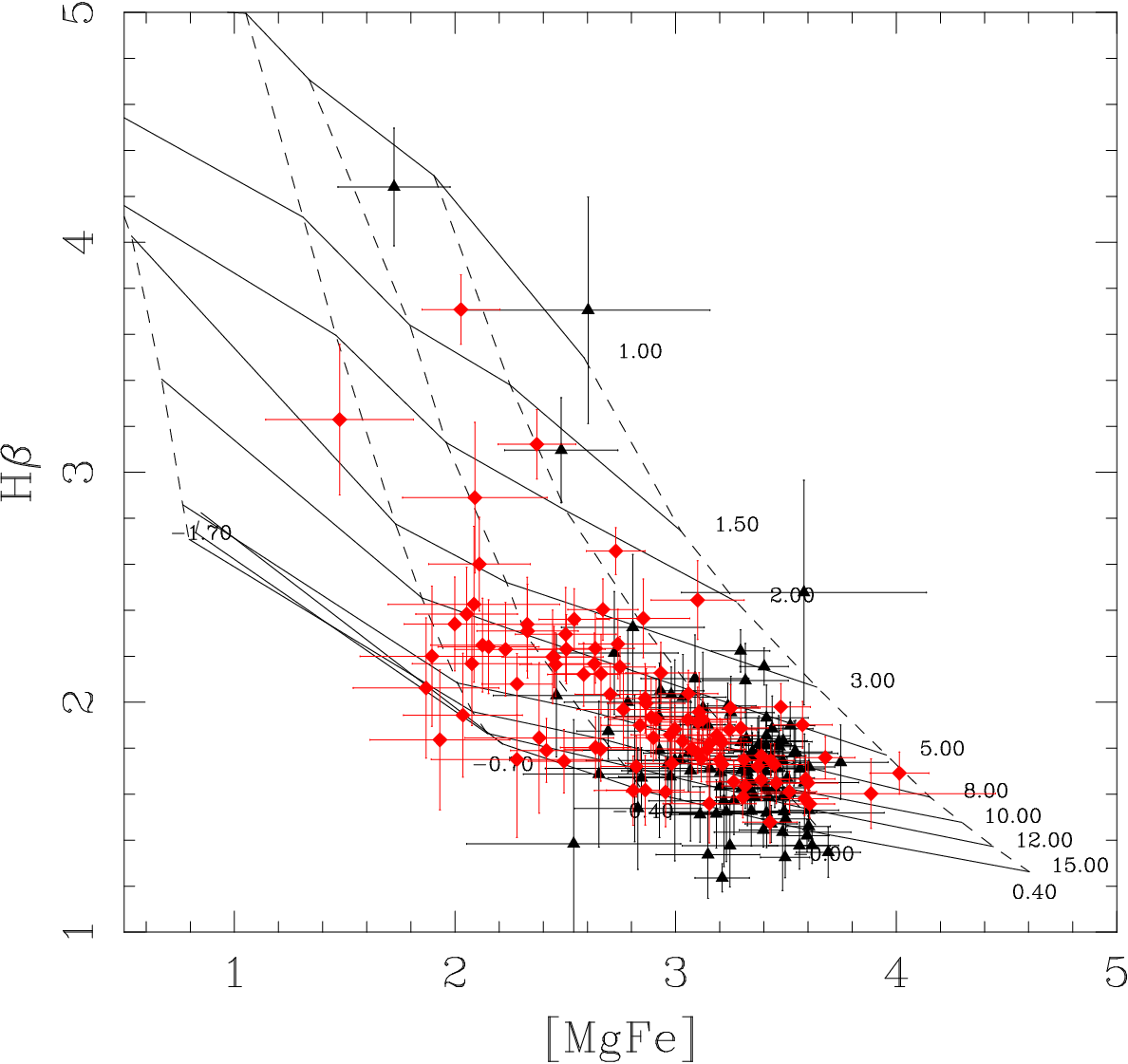}
  \caption{As for Fig.~\ref{fig:scthbmgbfe}, but for the \citet{Nelan05}
  sample.\label{fig:nelanhbmgbfe}}
\end{figure}

\begin{figure}
  \includegraphics[width=85mm]{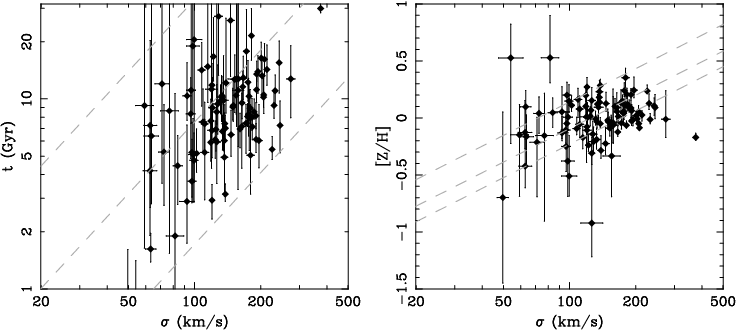}

  \includegraphics[width=85mm]{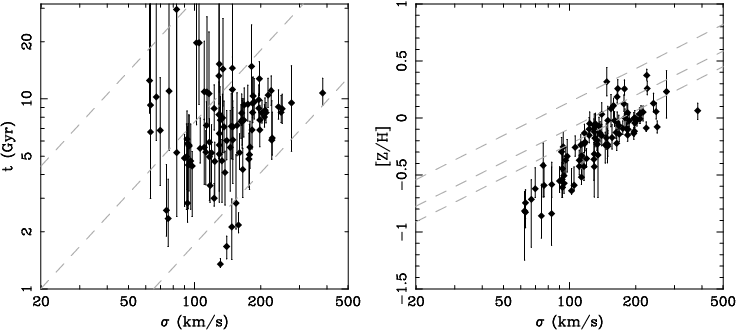}
  \caption{As for Fig.~\ref{fig:sctsigmarels}, but for the \citet{Nelan05}
  sample.\label{fig:nelansigmarels}}
\end{figure}

\begin{figure}
  \includegraphics[width=85mm]{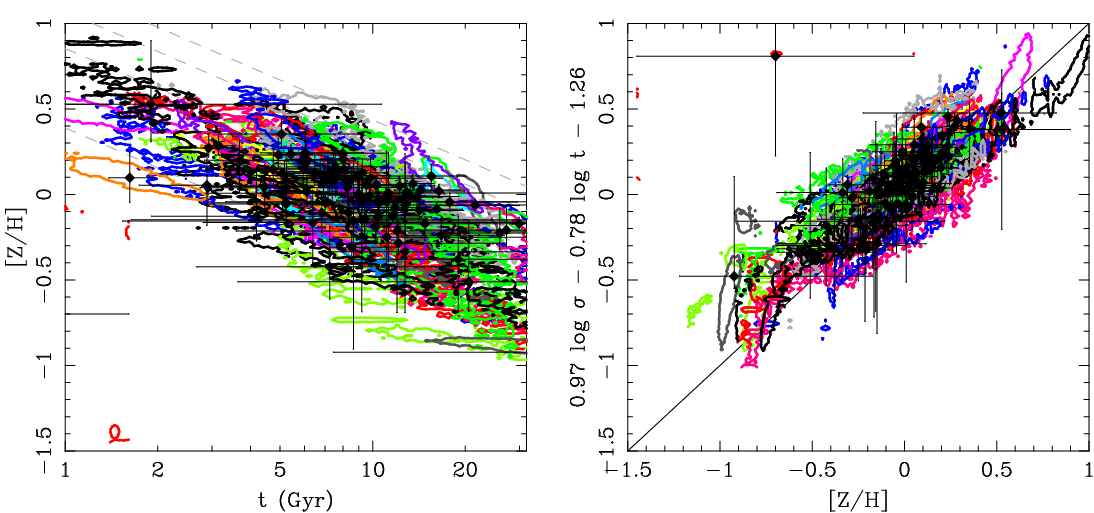}

  \includegraphics[width=85mm]{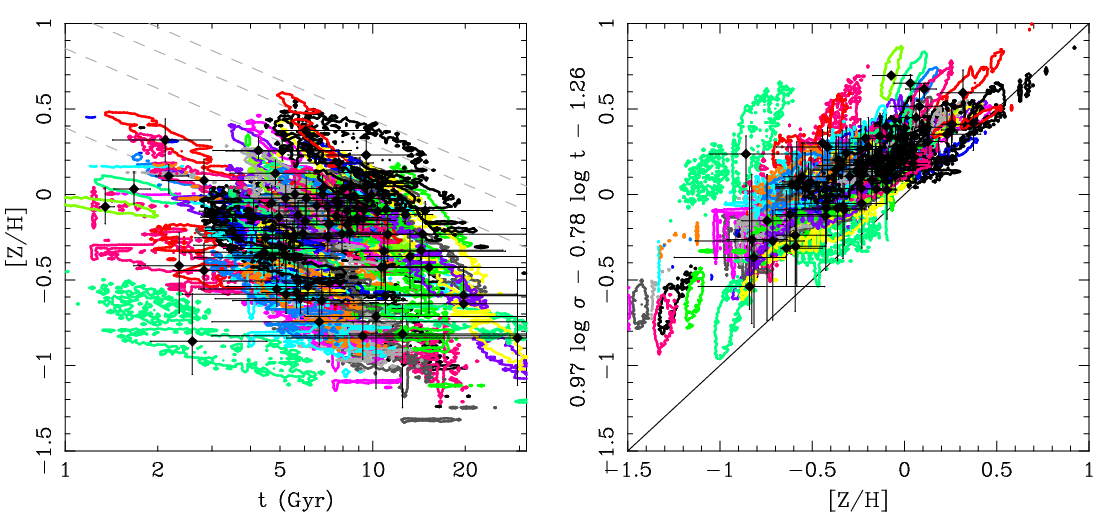}
  \caption{As for Fig.~\ref{fig:sctzplane}, but for the \citet{Nelan05}
  sample.\label{fig:nelanzplane}}
\end{figure}

We repeat this exercise with the Moore and Nelan samples in
Figures~\ref{fig:moorehbmgbfe}--\ref{fig:nelanzplane}.  It is clear
from these figures that the poorer data quality of the Moore and Nelan
samples makes definitive conclusions difficult, but the same trends
seen in the LRIS sample are generally present.  Although the
lowest-velocity dispersion model galaxies are missing due to the
simulation mass limit (see above), there seems to be a lack of model
galaxies with ages $\tssp<4$ Gyr compared with the \citet{M02} sample
at $\sigma\la200\,\kms$ in the left-hand panels of
Figure~\ref{fig:mooresigmarels}; the effect is less severe for the
\citet{Nelan05} sample (left-hand panels of
Fig.~\ref{fig:nelansigmarels}).  However, the errors in the
observations are large enough to make a quantification of this result
problematic.

It is also clear from the right-hand panels of
Figures~\ref{fig:mooresigmarels} and \ref{fig:nelansigmarels} that the
mass--metallicity relations of the mock catalogues are steeper than
those of the real data, as the large number of galaxies makes
comparison of the relations easier than in
Figure~\ref{fig:sctsigmarels}, although this result is also apparent
there.  We again see an offset in the metallicities in these figures
larger than the likely aperture effect.  Using galaxy formation models
with more detailed chemical evolution prescriptions by Arrigoni et
al.\ (in prep.), we have found that the slope difference in the
mass--metallicity relation is mostly corrected by the proper inclusion
of SNe Ia in the chemical evolution model, and that the metallicity
offset can be corrected by a slight flattening of the initial mass
function.  However, proper inclusion of this more detailed modelling
here is beyond the scope of the current paper.  The apparent smaller
scatter in the metallicity ditributions of the simulated galaxies at
low velocity dispersion in the lower right-hand panels of
Figures~\ref{fig:mooresigmarels} and \ref{fig:nelansigmarels} is
interesting, as the scatter in the data appears larger.  It is likely
that much of this scatter in the data is from observational
uncertainties (e.g., poor emission-line correction and low
signal-to-noise), but that may not explain all of the scatter, and the
tightness of the model metallicities may be another problem in the
current models.  It would be interesting to attempt to reproduce the
recent results of \citet{Smith08}, who studied dwarf galaxies at high
signal-to-noise in the Coma cluster, to pursue this further.

We conclude from the comparison of these three surveys of Coma
galaxies with their mock catalogues that the current semi-analytic
models reproduce the present-day SSP-equivalent ages moderately well,
although the agreement is not perfect.  The model apparently does not
have the correct star-formation histories for these objects, such that
the `archaeological downsizing' seen in model galaxies is stronger
than in the best observations.  However, the moderately good agreement
between the SSP-equivalent ages of the mock catalogues and the data
suggests that the star-formation quenching recipe implemented in the
hierarchical galaxy formation model produces roughly the correct
amount of recent star formation and is therefore a reasonable if
imperfect representation of the process suppressing star formation in
these galaxies.  The model metallicities are clearly problematic due
in large part to known deficiencies with the simple chemical evolution
prescription used here.  We expect to correct these deficiencies in
upcoming models (Arrigoni et al., in prep.).

\subsection{Idealised nearby rich cluster surveys}
\label{sec:ideal}

\begin{figure}
\includegraphics[width=85mm]{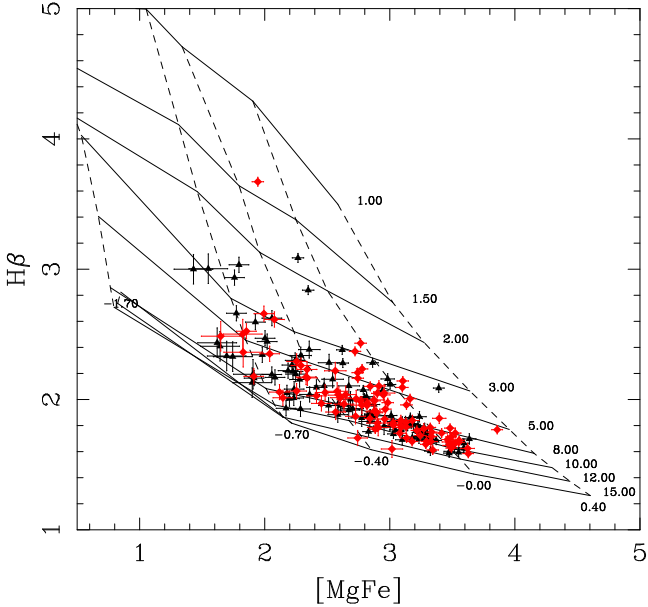}
\caption{The absorption-line strengths of two idealised nearby rich
  cluster galaxy surveys.  Black points are model galaxies selected to
  have early-type morphology; red points are model galaxies selected
  to be on the red sequence (with the same limits as the
  \citealt{Nelan05} sample).  Model grids as in
  Fig.~\ref{fig:hbmgbfe}.  Both mock catalogues have the same
  line-strength errors as the TFD08 sample; the morphology-selected
  sample has the same number of galaxies and VDF as the \citet[see
    Fig.~\ref{fig:vdf}, bottom panel]{M02} sample; and the
  red-sequence-colour-selected sample has the same number of galaxies
  and VDF as the \citet{Nelan05} sample.  The age, metallicity, and
  enhancement ratio distributions of these samples can be easily
  determined with such a large sample with such small uncertainties.
  Nearly all young, metal-rich galaxies in the red-sequence sample are
  dusty late-type galaxies.\label{fig:idealhbmgbfe}}
\end{figure}

\begin{figure}
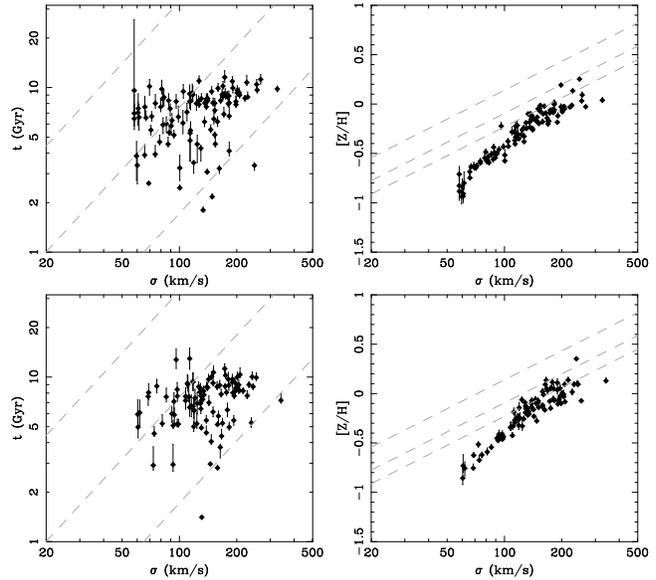

  \includegraphics[width=85mm]{tze_ideal_coma_sams_bc03_sf_stpops-sigmas.eps}

  \includegraphics[width=85mm]{tze_ideal2_coma_sams_bc03_sf_stpops-sigmas.eps}
  \caption{The predicted velocity dispersion--stellar population
    relations for two idealised nearby rich cluster surveys.  Top
    panel: morphology-selected sample.  Points with large error bars
    in age (right panel) are those that sit very close to or off the
    bottom of the SSP model grids in Fig.~\ref{fig:idealhbmgbfe}.
    Bottom panel: red-sequence-colour-selected sample.  As in
    Fig.~\ref{fig:idealhbmgbfe}, the age, metallicity, and enhancement
    ratio distributions can be easily measured from such surveys, but
    the `archaeological downsizing' seen in mass-weighted age in
    Fig.~\ref{fig:agemass} is difficult to see even in this
    sample.\label{fig:idealsigmarels}}
\end{figure}

\begin{figure}
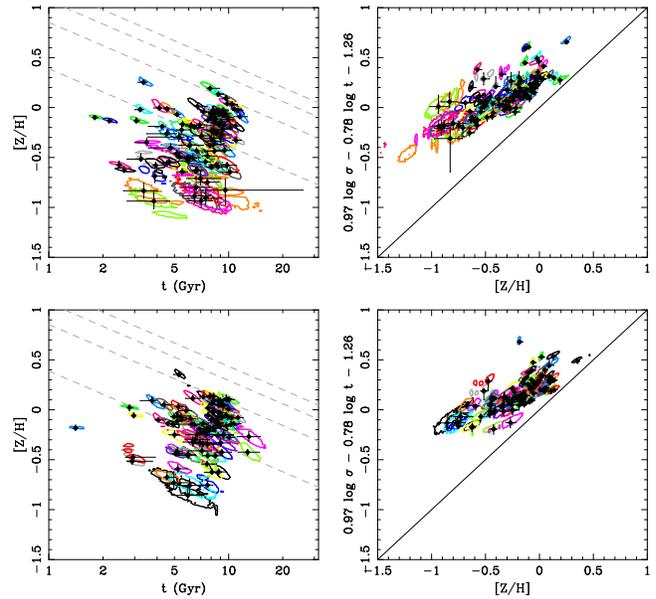

  \includegraphics[width=85mm]{tze_ideal_coma_sams_bc03_sf_stpops-zplane.eps}

  \includegraphics[width=85mm]{tze_ideal2_coma_sams_bc03_sf_stpops-zplane.eps}
  \caption{As in Figs.~\ref{fig:moorezplane} and \ref{fig:sctzplane},
    for two idealised nearby rich cluster surveys.  Top panel:
    morphology-selected sample.  Bottom panel:
    red-sequence-colour-selected sample.  In comparison with
    Fig.~\ref{fig:moorezplane}, the error contours are small enough to
    see clear deviations from the plane, as well as any curvature that
    may be present.\label{fig:idealzplane}}
\end{figure}

In the previous subsection we found that the currently available
samples of Coma cluster galaxies are either too small or of too low
quality to provide sharp tests of the hierarchical galaxy formation
models.  We now ask what could be learned from a spectroscopic survey
of a Coma-sized cluster with roughly 100 galaxies (similar to the
\citealt{M02} and \citealt{Nelan05} samples) with the data quality of
the LRIS (TFD08) sample.  

We create two different idealised large surveys: one selected solely
by morphology to be early-type galaxies with any colour and one solely
by colour to lie on the red sequence with any morphology (although
both are mass-limited by the simulation itself).  To be precise, we
selected for the first sample early-type galaxies from the simulations
to have the VDF of the \citet{M02} sample and for the second the
red-sequence colour--magnitude selection function and VDF of
\citet{Smith04}.  In both cases we applied the line-strength
uncertainties of the galaxy in the LRIS sample with velocity
dispersion nearest to each model galaxy to the model line strengths.

We plot the resulting line-strengths in Figure~\ref{fig:idealhbmgbfe}
and SSP-equivalent parameters as functions of velocity dispersion in
Figure~\ref{fig:idealsigmarels}.  Note that the early-type galaxy
sample has more young ($\tssp<5$) galaxies than the red-sequence
sample; these are blue early-type galaxies, absent from the latter
sample.  Nearly all of the young, metal-rich galaxies in the
red-sequence sample are dusty, late-type (spiral) galaxies in the
simulation.  On the other hand, it appears that there may be more
low-mass, old but metal-poor galaxies in the early-type galaxy sample
than in the red-sequence sample.  This is apparently a result of the
magnitude limit of the latter sample, which selects against the
smallest and therefore metal-poorest galaxies; however, this is a
small effect, as the magnitude limit of the red-sequence selection
($R_c<17$) is close to the mass limit in our simulation for old
galaxies at this redshift (Fig.~\ref{fig:cmd}).

It is clear from Figures~\ref{fig:idealhbmgbfe}--\ref{fig:idealzplane}
that a large, high-quality survey of a Coma-sized cluster would
provide a wealth of information on the precise distribution of stellar
population parameters in early-type and/or red-sequence cluster
galaxies, showing any deviations from or curvature of the $Z$ plane
and providing stringent tests of both hierarchical galaxy formation
models and previous data.  As a simple but interesting example, such a
survey would answer the question posed by TFD08 \citep[see
also][]{SB06a,SB06b}: are the nearly constant ages of early-type Coma
cluster galaxies just a selection effect resulting from observing only
12 galaxies, or is this a general result for the entire cluster?

\section{Conclusions}
\label{sec:conclusions}

In this study we set out to use hierarchical galaxy formation models
to understand the properties of local galaxies and, conversely, to use
those properties to test hierarchical galaxy formation models.  To
accomplish these goals, we make use of a state-of-the-art
semi-analytic galaxy formation model that includes a well-calibrated
parametrization of galaxy mergers, black hole growth and both `bright'
and `radio mode' feedback from these black holes. We coupled these
cosmologically motivated models to stellar population models to
produce synthetic spectra and computed line-strengths from these
spectra.  We then used the line strengths to determine `single-stellar
population-equivalent' (SSP-equivalent) ages and compositions \emph{in
precisely the same way as for the observed spectra}.

We created 20 realisations of a simulated Coma cluster-sized halo for
this study in order to compare our results with recent studies of
early-type galaxies in the Coma cluster \citep{M02,Nelan05,TFD08}.  We
began by comparing the inferred SSP-equivalent ages and compositions
of \emph{model} early-type galaxies with the mass- and light-weighted
ages and compositions of these model galaxies.  We found that
SSP-equivalent ages are (at best) biased tracers of mass- or
light-weighted ages, and are always younger than than these `true'
ages.  This is because hot, young stars contribute much more strongly
per unit mass to the Balmer lines than old stars
\citep{T00b,TWFD05,ST07}.  The bias of SSP-equivalent ages towards
young ages implies that the `archaeological' or `fossil downsizing'
seen in previous studies \citep[e.g.,][]{TMBO05,Nelan05,Clemens06}
overstates the `true' downsizing in the early-type galaxy population.
Rather, we find that the SSP-equivalent age correlates most strongly
with the fraction of stars formed within the past Gyr.  On the other
hand, the SSP-equivalent metallicity is an excellent tracer of the
light-weighted metallicity and a very good tracer of the mass-weighted
metallicity of a galaxy.  This is because hot, young stars contribute
little to the metal lines in a composite spectrum \citep{TWFD05,ST07}.
These conclusions strengthen previous suggestions about the behaviour
of line-strength-derived ages and metallicities based on simple
two-burst stellar population models but extend them to more complex
and, presumably, more realistic star-formation histories from a
fully-cosmological hierarchical galaxy formation model.

We then extracted mock catalogues from the simulation to represent
three recent line-strength surveys of early-type galaxies in the Coma
cluster: \citet{M02}, \citet{Nelan05}, and \citet{TFD08}.  These mock
catalogues were intended to help us both to interpret the observations
and to test the galaxy formation model itself.  Unfortunately, drawing
significant conclusions about the star-formation histories of Coma
cluster galaxies was not possible: the size of the high-quality sample
(TFD08) was too small, while the data quality of the large samples
\citep{M02,Nelan05} was too poor.  We could however note some
interesting problems with the model from discrepancies between the
mock catalogues and the observed samples.  First, comparison of the
observed velocity-dispersion--metallicity relations with the models
revealed that the models are currently producing neither the correct
zero-point of these relations -- the model galaxies are too metal-poor
-- nor the correct slope.  The former problem is likely to be due to
the lack of SNe Ia nucleosynthesis in the current models, while the
latter problem has two likely causes: (1) the star-formation histories
of the model galaxies are not the same as real galaxies, such that the
variation of \tssp\ with velocity dispersion is too steep in the
models compared with the data (left-hand panels of
Fig.~\ref{fig:sctsigmarels}), leading to an incorrect
\tssp--\zssp\ relation (left-hand panels of Fig.~\ref{fig:sctzplane});
and (2) the assumed IMF may not be correct (Arrigoni et al., in
prep.).  That being said, the reasonable agreement between the
SSP-equivalent ages of the mock catalogues and the observed samples
suggests that, while imperfect, the star-formation quenching model
implemented in the hierarchical galaxy formation models is a
reasonable representation of the true process suppressing star
formation in early-type galaxies.  We suggest that a large, deep,
high-quality survey of a cluster like Coma -- easily achievable with
the current generation of wide-field spectrographs on 8--10-metre
class telescopes -- would provide a sharp test for the models that is
not yet possible with the current data sets.

We have shown that the combination of hierarchical galaxy formation
models with detailed stellar population models provides powerful tools
both for understanding the stellar populations of early-type galaxies
and for testing the hierarchical galaxy formation models themselves.
We look forward to better datasets and to more realistic models for
both galaxy formation and stellar populations.  As has been mentioned
several times, we are currently working on one of these aspects,
introducing a detailed treatment of multi-element chemical evolution
into the semi-analytic galaxy formation models (Arrigoni et
al., in prep.).  These enhanced models will allow us to probe the
star-formation histories of galaxies -- and test the galaxy formation
models -- much more robustly and thoroughly than previously possible.

\section*{Acknowledgments}

We would like to thank Sandy Faber for suggesting the combination of
semi-analytic and stellar population models that inspired this study.
Peter Polko developed an early version of these models, for which we
are grateful.  We also thank Mat{\'\i}as Arrigoni for a careful
reading of the manuscript, Eric Bell, Richard Bower and Paolo Serra
for stimulating conversations, and the anonymous referee for
suggestions that helped to clarify the presentation.  Finally, we
thank the directors of the Max-Planck-Institut f\"ur Astronomie,
Hans-Walter Rix, and the Kapteyn Astronomical Institute, J.M.~van der
Hulst, and the Leids Kerkhoven-Bosscha Fonds for providing travel
support and working space during the gestation of this paper.

\bibliographystyle{mn2e}
\bibliography{models}

\begin{thebibliography}{}

\bibitem[\protect\citeauthoryear{{Adelman-McCarthy} et~al.,}{{Adelman-McCarthy}
   et~al.}{2008}]{DR6}
{Adelman-McCarthy} J.~K.,  et~al., 2008, ApJS, 175, 297

\bibitem[\protect\citeauthoryear{{Baldry}, {Glazebrook}, {Brinkmann},
  {Ivezi{\'c}}, {Lupton}, {Nichol} \& {Szalay}}{{Baldry}
  et~al.}{2004}]{Baldry04}
{Baldry} I.~K.,  {Glazebrook} K.,  {Brinkmann} J.,  {Ivezi{\'c}} {\v Z}.,
  {Lupton} R.~H.,  {Nichol} R.~C.,    {Szalay} A.~S.,  2004, ApJ, 600, 681

\bibitem[\protect\citeauthoryear{{Bernardi}, {Sheth}, {Nichol}, {Schneider} \&
  {Brinkmann}}{{Bernardi} et~al.}{2005}]{Bernardi05}
{Bernardi} M.,  {Sheth} R.~K.,  {Nichol} R.~C.,  {Schneider} D.~P.,
  {Brinkmann} J.,  2005, AJ, 129, 61

\bibitem[\protect\citeauthoryear{{Bernardi}, {Nichol}, {Sheth}, {Miller} \&
  {Brinkmann}}{{Bernardi} et~al.}{2006}]{Bernardi06}
{Bernardi} M.,  {Nichol} R.~C.,  {Sheth} R.~K.,  {Miller} C.~J.,    {Brinkmann}
  J.,  2006, AJ, 131, 1288

\bibitem[\protect\citeauthoryear{{Bertelli}, {Bressan}, {Chiosi}, {Fagotto} \&
  {Nasi}}{{Bertelli} et~al.}{1994}]{Padova}
{Bertelli} G.,  {Bressan} A.,  {Chiosi} C.,  {Fagotto} F.,    {Nasi} E.,  1994,
  A\&AS, 106, 275

\bibitem[\protect\citeauthoryear{{Blanton} \& {Roweis}}{{Blanton} \&
  {Roweis}}{2007}]{BR07}
{Blanton} M.~R.,  {Roweis} S.,  2007, AJ, 133, 734

\bibitem[\protect\citeauthoryear{{Blumenthal}, {Faber}, {Primack} \&
  {Rees}}{{Blumenthal} et~al.}{1984}]{BFPR84}
{Blumenthal} G.~R.,  {Faber} S.~M.,  {Primack} J.~R.,    {Rees} M.~J.,  1984,
  Nature, 311, 517

\bibitem[\protect\citeauthoryear{{Bower}, {Benson}, {Malbon}, {Helly}, {Frenk},
  {Baugh}, {Cole} \& {Lacey}}{{Bower} et~al.}{2006}]{Bower06}
{Bower} R.~G.,  {Benson} A.~J.,  {Malbon} R.,  {Helly} J.~C.,  {Frenk} C.~S.,
  {Baugh} C.~M.,  {Cole} S.,    {Lacey} C.~G.,  2006, MNRAS, 370, 645

\bibitem[\protect\citeauthoryear{{Bruzual} \& {Charlot}}{{Bruzual} \&
  {Charlot}}{2003}]{BC03}
{Bruzual} G.,  {Charlot} S.,  2003, MNRAS, 344, 1000

\bibitem[\protect\citeauthoryear{{Buzzoni}, {Mantegazza} \&
  {Gariboldi}}{{Buzzoni} et~al.}{1994}]{Buzzoni94}
{Buzzoni} A.,  {Mantegazza} L.,    {Gariboldi} G.,  1994, AJ, 107, 513

\bibitem[\protect\citeauthoryear{Cardelli, Clayton \& Mathis}{Cardelli
  et~al.}{1989}]{cardelli:89}
Cardelli J.,  Clayton G.,    Mathis J.,  1989, ApJ, 345, 245

\bibitem[\protect\citeauthoryear{{Cardiel}, {Gorgas}, {Cenarro} \&
  {Gonzalez}}{{Cardiel} et~al.}{1998}]{Cardiel98}
{Cardiel} N.,  {Gorgas} J.,  {Cenarro} J.,    {Gonzalez} J.~J.,  1998, A\&AS,
  127, 597

\bibitem[\protect\citeauthoryear{{Cardiel}, {Gorgas},
  {S{\'a}nchez-Bl{\'a}zquez}, {Cenarro}, {Pedraz}, {Bruzual} \&
  {Klement}}{{Cardiel} et~al.}{2003}]{Cardiel03}
{Cardiel} N.,  {Gorgas} J.,  {S{\'a}nchez-Bl{\'a}zquez} P.,  {Cenarro} A.~J.,
  {Pedraz} S.,  {Bruzual} G.,    {Klement} J.,  2003, A\&A, 409, 511

\bibitem[\protect\citeauthoryear{{Cattaneo}, {Dekel}, {Devriendt}, {Guiderdoni}
  \& {Blaizot}}{{Cattaneo} et~al.}{2006}]{Cattaneo06}
{Cattaneo} A.,  {Dekel} A.,  {Devriendt} J.,  {Guiderdoni} B.,    {Blaizot} J.,
   2006, MNRAS, 370, 1651

\bibitem[\protect\citeauthoryear{{Chabrier}}{{Chabrier}}{2001}]{Chabrier01}
{Chabrier} G.,  2001, ApJ, 554, 1274

\bibitem[\protect\citeauthoryear{{Charlot} \& {Fall}}{{Charlot} \&
  {Fall}}{2000}]{charlot_fall:00}
{Charlot} S.,  {Fall} S.~M.,  2000, ApJ, 539, 718

\bibitem[\protect\citeauthoryear{{Clemens}, {Bressan}, {Nikolic}, {Alexander},
  {Annibali} \& {Rampazzo}}{{Clemens} et~al.}{2006}]{Clemens06}
{Clemens} M.~S.,  {Bressan} A.,  {Nikolic} B.,  {Alexander} P.,  {Annibali} F.,
     {Rampazzo} R.,  2006, MNRAS, 370, 702

\bibitem[\protect\citeauthoryear{{Cole}, {Aragon-Salamanca}, {Frenk}, {Navarro}
  \& {Zepf}}{{Cole} et~al.}{1994}]{Cole94}
{Cole} S.,  {Aragon-Salamanca} A.,  {Frenk} C.~S.,  {Navarro} J.~F.,    {Zepf}
  S.~E.,  1994, MNRAS, 271, 781

\bibitem[\protect\citeauthoryear{{Cox}, {Jonsson}, {Somerville}, {Primack} \&
  {Dekel}}{{Cox} et~al.}{2008}]{cox:08}
{Cox} T.~J.,  {Jonsson} P.,  {Somerville} R.~S.,  {Primack} J.~R.,    {Dekel}
  A.,  2008, MNRAS, 384, 386

\bibitem[\protect\citeauthoryear{{Croton}, {Springel}, {White}, {De Lucia},
  {Frenk}, {Gao}, {Jenkins}, {Kauffmann}, {Navarro} \& {Yoshida}}{{Croton}
  et~al.}{2006}]{Croton06}
{Croton} D.~J.,  {Springel} V.,  {White} S.~D.~M.,  {De Lucia} G.,  {Frenk}
  C.~S.,  {Gao} L.,  {Jenkins} A.,  {Kauffmann} G.,  {Navarro} J.~F.,
  {Yoshida} N.,  2006, MNRAS, 365, 11

\bibitem[\protect\citeauthoryear{{Davies}, {Sadler} \& {Peletier}}{{Davies}
  et~al.}{1993}]{DSP93}
{Davies} R.~L.,  {Sadler} E.~M.,    {Peletier} R.~F.,  1993, MNRAS, 262, 650

\bibitem[\protect\citeauthoryear{{De Lucia} \& {Blaizot}}{{De Lucia} \&
  {Blaizot}}{2007}]{delucia_blaizot:07}
{De Lucia} G.,  {Blaizot} J.,  2007, MNRAS, 375, 2

\bibitem[\protect\citeauthoryear{{De Lucia}, {Springel}, {White}, {Croton} \&
  {Kauffmann}}{{De Lucia} et~al.}{2006}]{deLucia06}
{De Lucia} G.,  {Springel} V.,  {White} S.~D.~M.,  {Croton} D.,    {Kauffmann}
  G.,  2006, MNRAS, 366, 499

\bibitem[\protect\citeauthoryear{{Di Matteo}, {Springel} \& {Hernquist}}{{Di
  Matteo} et~al.}{2005}]{dimatteo:05}
{Di Matteo} T.,  {Springel} V.,    {Hernquist} L.,  2005, Nature, 433, 604

\bibitem[\protect\citeauthoryear{{Faber}}{{Faber}}{1973}]{Faber73}
{Faber} S.~M.,  1973, ApJ, 179, 731

\bibitem[\protect\citeauthoryear{{Gallazzi}, {Charlot}, {Brinchmann}, {White}
  \& {Tremonti}}{{Gallazzi} et~al.}{2005}]{Gallazzi05}
{Gallazzi} A.,  {Charlot} S.,  {Brinchmann} J.,  {White} S.~D.~M.,
  {Tremonti} C.~A.,  2005, MNRAS, 362, 41

\bibitem[\protect\citeauthoryear{{Gonz{\'a}lez}}{{Gonz{\'a}lez}}{1993}]{G93}
{Gonz{\'a}lez} J.~J.,  1993, PhD thesis, University of California, Santa Cruz

\bibitem[\protect\citeauthoryear{{Graves}, {Faber}, {Schiavon} \&
  {Yan}}{{Graves} et~al.}{2007}]{Graves07}
{Graves} G.~J.,  {Faber} S.~M.,  {Schiavon} R.~P.,    {Yan} R.,  2007, ApJ,
  671, 243

\bibitem[\protect\citeauthoryear{{H{\"a}ring} \& {Rix}}{{H{\"a}ring} \&
  {Rix}}{2004}]{HR04}
{H{\"a}ring} N.,  {Rix} H.-W.,  2004, ApJL, 604, L89

\bibitem[\protect\citeauthoryear{{Heavens}, {Jimenez} \& {Lahav}}{{Heavens}
  et~al.}{2000}]{MOPED1}
{Heavens} A.~F.,  {Jimenez} R.,    {Lahav} O.,  2000, MNRAS, 317, 965

\bibitem[\protect\citeauthoryear{{Hopkins}, {Hernquist}, {Cox}, {Robertson} \&
  {Krause}}{{Hopkins} et~al.}{2007}]{hopkins_bhfpth:07}
{Hopkins} P.~F.,  {Hernquist} L.,  {Cox} T.~J.,  {Robertson} B.,    {Krause}
  E.,  2007, ApJ, 669, 45

\bibitem[\protect\citeauthoryear{{Hudson}, {Lucey}, {Smith}, {Schlegel} \&
  {Davies}}{{Hudson} et~al.}{2001}]{Hudson01}
{Hudson} M.~J.,  {Lucey} J.~R.,  {Smith} R.~J.,  {Schlegel} D.~J.,    {Davies}
  R.~L.,  2001, MNRAS, 327, 265

\bibitem[\protect\citeauthoryear{{Jimenez}, {Bernardi}, {Haiman}, {Panter} \&
  {Heavens}}{{Jimenez} et~al.}{2007}]{Jimenez07}
{Jimenez} R.,  {Bernardi} M.,  {Haiman} Z.,  {Panter} B.,    {Heavens} A.~F.,
  2007, ApJ, 669, 947

\bibitem[\protect\citeauthoryear{{Kauffmann}}{{Kauffmann}}{1996}]{Kauffmann96}
{Kauffmann} G.,  1996, MNRAS, 281, 487

\bibitem[\protect\citeauthoryear{{Kauffmann} \& {Charlot}}{{Kauffmann} \&
  {Charlot}}{1998}]{KC98}
{Kauffmann} G.,  {Charlot} S.,  1998, MNRAS, 294, 705

\bibitem[\protect\citeauthoryear{{Kauffmann}, {White} \&
  {Guiderdoni}}{{Kauffmann} et~al.}{1993}]{KWG93}
{Kauffmann} G.,  {White} S.~D.~M.,    {Guiderdoni} B.,  1993, MNRAS, 264, 201

\bibitem[\protect\citeauthoryear{Kennicutt}{Kennicutt}{1989}]{kennicutt:89}
Kennicutt R.,  1989, ApJ, 344, 685

\bibitem[\protect\citeauthoryear{Kennicutt}{Kennicutt}{1998}]{kennicutt:98}
Kennicutt R.,  1998, ApJ, 498, 181

\bibitem[\protect\citeauthoryear{{Kubo}, {Stebbins}, {Annis}, {Dell'Antonio},
  {Lin}, {Khiabanian} \& {Frieman}}{{Kubo} et~al.}{2007}]{Kubo07}
{Kubo} J.~M.,  {Stebbins} A.,  {Annis} J.,  {Dell'Antonio} I.~P.,  {Lin} H.,
  {Khiabanian} H.,    {Frieman} J.~A.,  2007, ApJ, 671, 1466

\bibitem[\protect\citeauthoryear{{Kuntschner}}{{Kuntschner}}{2000}]{Kuntschner%
00}
{Kuntschner} H.,  2000, MNRAS, 315, 184

\bibitem[\protect\citeauthoryear{{Kuntschner}, {Lucey}, {Smith}, {Hudson} \&
  {Davies}}{{Kuntschner} et~al.}{2001}]{Kuntschner01}
{Kuntschner} H.,  {Lucey} J.~R.,  {Smith} R.~J.,  {Hudson} M.~J.,    {Davies}
  R.~L.,  2001, MNRAS, 323, 615

\bibitem[\protect\citeauthoryear{{Lintott}, {Ferreras} \& {Lahav}}{{Lintott}
  et~al.}{2006}]{LFL06}
{Lintott} C.~J.,  {Ferreras} I.,    {Lahav} O.,  2006, ApJ, 648, 826

\bibitem[\protect\citeauthoryear{{{\L}okas} \& {Mamon}}{{{\L}okas} \&
  {Mamon}}{2003}]{LM03}
{{\L}okas} E.~L.,  {Mamon} G.~A.,  2003, MNRAS, 343, 401

\bibitem[\protect\citeauthoryear{{MacArthur}}{{MacArthur}}{2005}]{MacArthur05}
{MacArthur} L.~A.,  2005, ApJ, 623, 795

\bibitem[\protect\citeauthoryear{{Mehlert}, {Saglia}, {Bender} \&
  {Wegner}}{{Mehlert} et~al.}{2000}]{Mehlert00}
{Mehlert} D.,  {Saglia} R.~P.,  {Bender} R.,    {Wegner} G.,  2000, A\&AS, 141,
  449

\bibitem[\protect\citeauthoryear{{Moore}, {Lucey}, {Kuntschner} \&
  {Colless}}{{Moore} et~al.}{2002}]{M02}
{Moore} S.~A.~W.,  {Lucey} J.~R.,  {Kuntschner} H.,    {Colless} M.,  2002,
  MNRAS, 336, 382

\bibitem[\protect\citeauthoryear{{Nelan}, {Smith}, {Hudson}, {Wegner}, {Lucey},
  {Moore}, {Quinney} \& {Suntzeff}}{{Nelan} et~al.}{2005}]{Nelan05}
{Nelan} J.~E.,  {Smith} R.~J.,  {Hudson} M.~J.,  {Wegner} G.~A.,  {Lucey}
  J.~R.,  {Moore} S.~A.~W.,  {Quinney} S.~J.,    {Suntzeff} N.~B.,  2005, ApJ,
  632, 137

\bibitem[\protect\citeauthoryear{{Nulsen} \& {Fabian}}{{Nulsen} \&
  {Fabian}}{2000}]{nulsen_fabian:00}
{Nulsen} P.~E.~J.,  {Fabian} A.~C.,  2000, MNRAS, 311, 346

\bibitem[\protect\citeauthoryear{{O'Connell}}{{O'Connell}}{1986}]{OConnell86}
{O'Connell} R.~W.,  1986, in {Norman} C.~A.,  {Renzini} A.,   {Tosi} M.,  eds,
  Stellar Populations, Cambridge U. Press, Cambridge, p. 167

\bibitem[\protect\citeauthoryear{{Ocvirk}, {Pichon}, {Lan{\c c}on} \&
  {Thi{\'e}baut}}{{Ocvirk} et~al.}{2006}]{Ocvirk06a}
{Ocvirk} P.,  {Pichon} C.,  {Lan{\c c}on} A.,    {Thi{\'e}baut} E.,  2006,
  MNRAS, 365, 46

\bibitem[\protect\citeauthoryear{{Panter}, {Heavens} \& {Jimenez}}{{Panter}
  et~al.}{2003}]{Panter03}
{Panter} B.,  {Heavens} A.~F.,    {Jimenez} R.,  2003, MNRAS, 343, 1145

\bibitem[\protect\citeauthoryear{{Rabin}}{{Rabin}}{1982}]{Rabin82}
{Rabin} D.,  1982, ApJ, 261, 85

\bibitem[\protect\citeauthoryear{{Reichardt}, {Jimenez} \&
  {Heavens}}{{Reichardt} et~al.}{2001}]{MOPED2}
{Reichardt} C.,  {Jimenez} R.,    {Heavens} A.~F.,  2001, MNRAS, 327, 849

\bibitem[\protect\citeauthoryear{{Renzini}}{{Renzini}}{2006}]{Renzini06}
{Renzini} A.,  2006, ARA\&A, 44, 141

\bibitem[\protect\citeauthoryear{{Robertson}, {Cox}, {Hernquist}, {Franx},
  {Hopkins}, {Martini} \& {Springel}}{{Robertson} et~al.}{2006}]{robertson:06}
{Robertson} B.,  {Cox} T.~J.,  {Hernquist} L.,  {Franx} M.,  {Hopkins} P.~F.,
  {Martini} P.,    {Springel} V.,  2006, ApJ, 641, 21

\bibitem[\protect\citeauthoryear{{S{\'a}nchez-Bl{\'a}zquez}, {Gorgas},
  {Cardiel} \& {Gonz{\'a}lez}}{{S{\'a}nchez-Bl{\'a}zquez}
  et~al.}{2006a}]{SB06a}
{S{\'a}nchez-Bl{\'a}zquez} P.,  {Gorgas} J.,  {Cardiel} N.,    {Gonz{\'a}lez}
  J.~J.,  2006a, A\&A, 457, 787

\bibitem[\protect\citeauthoryear{{S{\'a}nchez-Bl{\'a}zquez}, {Gorgas},
  {Cardiel} \& {Gonz{\'a}lez}}{{S{\'a}nchez-Bl{\'a}zquez}
  et~al.}{2006b}]{SB06b}
{S{\'a}nchez-Bl{\'a}zquez} P.,  {Gorgas} J.,  {Cardiel} N.,    {Gonz{\'a}lez}
  J.~J.,  2006b, A\&A, 457, 809

\bibitem[\protect\citeauthoryear{{Schiavon}}{{Schiavon}}{2007}]{Schiavon07}
{Schiavon} R.~P.,  2007, ApJS, 171, 146

\bibitem[\protect\citeauthoryear{{Serra} \& {Trager}}{{Serra} \&
  {Trager}}{2007}]{ST07}
{Serra} P.,  {Trager} S.~C.,  2007, MNRAS, 374, 769

\bibitem[\protect\citeauthoryear{{Simien} \& {de Vaucouleurs}}{{Simien} \& {de
  Vaucouleurs}}{1986}]{SdV86}
{Simien} F.,  {de Vaucouleurs} G.,  1986, ApJ, 302, 564

\bibitem[\protect\citeauthoryear{{Smith} et~al.,}{{Smith}
  et~al.}{2004}]{Smith04}
{Smith} R.~J.,  et~al., 2004, AJ, 128, 1558

\bibitem[\protect\citeauthoryear{{Smith}, {Lucey}, {Hudson}, {Allanson},
  {Bridges}, {Hornschemeier}, {Marzke} \& {Miller}}{{Smith}
  et~al.}{2008}]{Smith08}
{Smith} R.~J.,  {Lucey} J.~R.,  {Hudson} M.~J.,  {Allanson} S.~P.,  {Bridges}
  T.~J.,  {Hornschemeier} A.~E.,  {Marzke} R.~O.,    {Miller} N.~A.,  2008,
  ArXiv e-prints

\bibitem[\protect\citeauthoryear{{Somerville} \& {Kolatt}}{{Somerville} \&
  {Kolatt}}{1999}]{sk:99}
{Somerville} R.~S.,  {Kolatt} T.~S.,  1999, MNRAS, 305, 1

\bibitem[\protect\citeauthoryear{{Somerville} \& {Primack}}{{Somerville} \&
  {Primack}}{1999}]{sp}
{Somerville} R.~S.,  {Primack} J.~R.,  1999, MNRAS, 310, 1087

\bibitem[\protect\citeauthoryear{{Somerville}, {Primack} \&
  {Faber}}{{Somerville} et~al.}{2001}]{spf}
{Somerville} R.~S.,  {Primack} J.~R.,    {Faber} S.~M.,  2001, MNRAS, 320, 504

\bibitem[\protect\citeauthoryear{{Somerville} et~al.,}{{Somerville}
  et~al.}{2004}]{somerville:04}
{Somerville} R.~S.,  et~al., 2004, ApJL, 600, L135

\bibitem[\protect\citeauthoryear{{Somerville} et~al.,}{{Somerville}
  et~al.}{2008a}]{somerville:08b}
{Somerville} R.~S.,  et~al., 2008a, ApJ, 672, 776

\bibitem[\protect\citeauthoryear{{Somerville}, {Hopkins}, {Cox}, {Robertson},
  \& {Hernquist}}{{Somerville} et~al.}{2008b}]{s08}
{Somerville} R.~S.,  {Hopkins} P.~F.,  {Cox} T.~J.,  {Robertson} B.,
  {Hernquist} L.,  2008b, MNRAS, 391, 481

\bibitem[\protect\citeauthoryear{{Spergel} et~al.,}{{Spergel}
  et~al.}{2007}]{WMAP3}
{Spergel} D.~N.,  et~al., 2007, ApJS, 170, 377

\bibitem[\protect\citeauthoryear{{Springel}, {Di Matteo} \&
  {Hernquist}}{{Springel} et~al.}{2005}]{springel:05a}
{Springel} V.,  {Di Matteo} T.,    {Hernquist} L.,  2005, MNRAS, 361, 776

\bibitem[\protect\citeauthoryear{{Springel}, {White}, {Tormen} \&
  {Kauffmann}}{{Springel} et~al.}{2001}]{springel:01}
{Springel} V.,  {White} S.~D.~M.,  {Tormen} G.,    {Kauffmann} G.,  2001,
  MNRAS, 328, 726

\bibitem[\protect\citeauthoryear{{Strateva} et~al.,}{{Strateva}
  et~al.}{2001}]{Strateva01}
{Strateva} I.,  et~al., 2001, AJ, 122, 1861

\bibitem[\protect\citeauthoryear{{Thomas}, {Maraston} \& {Bender}}{{Thomas}
  et~al.}{2003}]{TMB03}
{Thomas} D.,  {Maraston} C.,    {Bender} R.,  2003, MNRAS, 339, 897

\bibitem[\protect\citeauthoryear{{Thomas}, {Maraston} \& {Korn}}{{Thomas}
  et~al.}{2004}]{TMK04}
{Thomas} D.,  {Maraston} C.,    {Korn} A.,  2004, MNRAS, 351, L19

\bibitem[\protect\citeauthoryear{{Thomas}, {Maraston}, {Bender} \& {Mendes de
  Oliveira}}{{Thomas} et~al.}{2005}]{TMBO05}
{Thomas} D.,  {Maraston} C.,  {Bender} R.,    {Mendes de Oliveira} C.,  2005,
  ApJ, 621, 673

\bibitem[\protect\citeauthoryear{{Thomas}, {Maraston}, {Schawinski}, {Sarzi},
  {Joo}, {Kaviraj} \& {Yi}}{{Thomas} et~al.}{2007}]{Thomas07}
{Thomas} D.,  {Maraston} C.,  {Schawinski} K.,  {Sarzi} M.,  {Joo} S.-J.,
  {Kaviraj} S.,    {Yi} S.~K.,  2007, in {Vazdekis} A.,  {Peletier} R.~F.,
  eds, proc. IAU Symp.\ 241, Stellar Populations as Building Blocks of
  Galaxies, Cambridge U. Press, Cambridge, p. 546

\bibitem[\protect\citeauthoryear{{Tojeiro}, {Heavens}, {Jimenez} \&
  {Panter}}{{Tojeiro} et~al.}{2007}]{VESPA}
{Tojeiro} R.,  {Heavens} A.~F.,  {Jimenez} R.,    {Panter} B.,  2007, MNRAS,
  381, 1252

\bibitem[\protect\citeauthoryear{{Toomre}}{{Toomre}}{1977}]{Toomre77}
{Toomre} A.,  1977, in {Tinsley} B.~M.,  {Larson} R.~B.,  eds, Evolution of
  Galaxies and Stellar Populations, Yale U. Press, New Haven, p. 401

\bibitem[\protect\citeauthoryear{{Trager}}{{Trager}}{1997}]{T97}
{Trager} S.~C.,  1997, PhD thesis, University of California, Santa Cruz

\bibitem[\protect\citeauthoryear{{Trager}, {Faber}, {Worthey} \&
  {Gonz{\'a}lez}}{{Trager} et~al.}{2000a}]{T00a}
{Trager} S.~C.,  {Faber} S.~M.,  {Worthey} G.,    {Gonz{\'a}lez} J.~J.,  2000a,
  AJ, 119, 1645

\bibitem[\protect\citeauthoryear{{Trager}, {Faber}, {Worthey} \&
  {Gonz{\'a}lez}}{{Trager} et~al.}{2000b}]{T00b}
{Trager} S.~C.,  {Faber} S.~M.,  {Worthey} G.,    {Gonz{\'a}lez} J.~J.,  2000b,
  AJ, 120, 165

\bibitem[\protect\citeauthoryear{{Trager}, {Worthey}, {Faber} \&
  {Dressler}}{{Trager} et~al.}{2005}]{TWFD05}
{Trager} S.~C.,  {Worthey} G.,  {Faber} S.~M.,    {Dressler} A.,  2005, MNRAS,
  362, 2

\bibitem[\protect\citeauthoryear{{Trager}, {Faber} \& {Dressler}}{{Trager}
  et~al.}{2008}]{TFD08}
{Trager} S.~C.,  {Faber} S.~M.,    {Dressler} A.,  2008, MNRAS, 386,
715 (TFD08)

\bibitem[\protect\citeauthoryear{{Vazdekis}, {Casuso}, {Peletier} \&
  {Beckman}}{{Vazdekis} et~al.}{1996}]{Vazdekis96}
{Vazdekis} A.,  {Casuso} E.,  {Peletier} R.~F.,    {Beckman} J.~E.,  1996,
  ApJS, 106, 307

\bibitem[\protect\citeauthoryear{{White} \& {Rees}}{{White} \&
  {Rees}}{1978}]{WR78}
{White} S.~D.~M.,  {Rees} M.~J.,  1978, MNRAS, 183, 341

\bibitem[\protect\citeauthoryear{{Worthey}}{{Worthey}}{1994}]{W94}
{Worthey} G.,  1994, ApJS, 95, 107

\bibitem[\protect\citeauthoryear{{Worthey} \& {Ottaviani}}{{Worthey} \&
  {Ottaviani}}{1997}]{WO97}
{Worthey} G.,  {Ottaviani} D.~L.,  1997, ApJS, 111, 377

\end{thebibliography}

\appendix
\section{Aperture corrections}

Unfortunately, the Coma cluster is not quite far enough away for its
member galaxies to be completely contained in a reasonably-sized slit
or fiber.  A fair comparison of the observational data with the galaxy
formation models requires corrections to be made for the fact that
early-type galaxies have line-strength gradients \citep[see,
  e.g.,][among many others]{DSP93,G93,Mehlert00}.  We must therefore
estimate the effect of these gradients on the stellar population
parameters.

\begin{figure}
  \includegraphics[width=85mm]{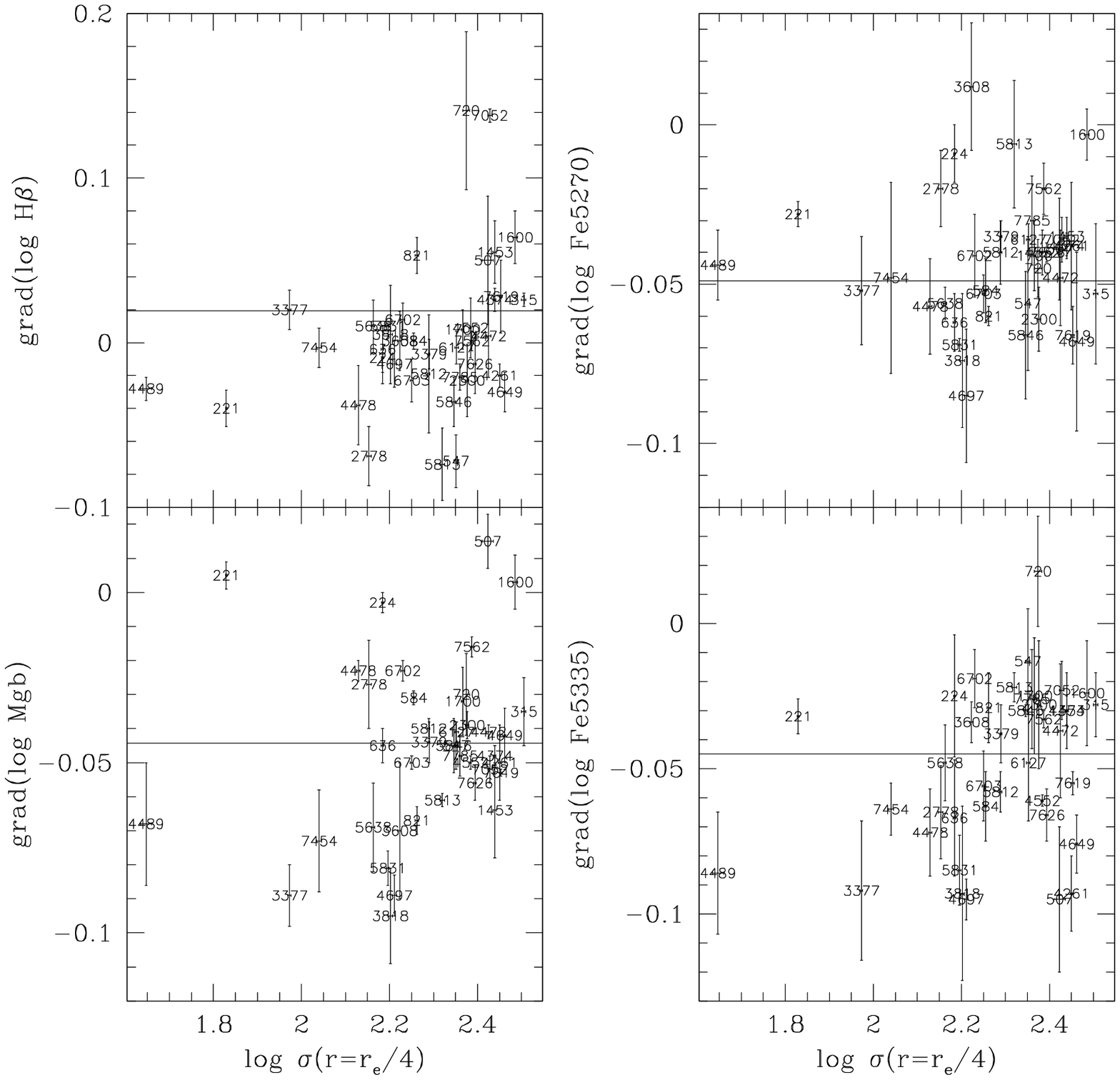}
  \caption{Line-strength gradients in elliptical apertures derived
    from \citet{G93}, in the sense $d\log I/d\log(r/r_e)$; see text
    for more details.  Solid lines are error-weighted means of the
    index gradients.}
  \label{fig:apgradients}
\end{figure}

To make the required corrections, we appeal to the high-quality
line-strength gradient measurements of \citet{G93}.  Gonz\'alez
measured gradients in 40 local early-type galaxies, typically along
both the major and minor axes.  We have used his
`elliptical-aperture-weighted' line-strengths (see his Chapter 5) and
taken error-weighted averages, where possible, of the major and minor
axis line strengths in apertures of radii $r=r_e/16$, $r_e/8$,
$r_e/4$, and $r_e/2$ (if only one axis was available, we used that
axis).  We then fit linear relations of the form
\begin{equation}
  \log I(r/r_e) = a + b \log(r/r_e),
\end{equation}
where $I(r/r_e)$ is the line strength within an elliptical aperture
with the fractional equivalent circular radius $r/r_e$.  We define
$b=d\log I(r/r_e)/d\log(r/r_e)$ the line-strength gradient for each
galaxy.  To search for trends in these gradients as a function of, for
example, velocity dispersion, we plot the results as a function of
$\log\sigma(r=r_e/4)$ in Figure~\ref{fig:apgradients}.  We find only
minimal hints of trends with velocity dispersion and a very large
scatter, so we use the error-weighted mean of the gradients from
\emph{all} galaxies for each index to estimate our aperture
corrections.  These means are tabulated in
Table~\ref{tbl:apgradients}.  The final (logarithmic) line strengths
at one effective radius -- a reasonable guess at a `global' value --
are then taken to be
\begin{equation}
  \log I(r_e) = \log I(r) + \langle b\rangle * \log(r/r_e),
\end{equation}
where $\langle b\rangle$ is the mean gradient of index $I$ and $r$ is
the radius of the original aperture: for the LRIS data, $=r_e/4$; for
the Moore sample, $r=1\farcs35$; and for the Nelan sample,
$r=1\farcs0$.  The errors in the mean gradients are added in
quadrature (after scaling by the logarithmic difference in the radii)
to the original line strength errors.  

\begin{table}
  \caption{Mean line-strength gradients derived from \citet{G93}}
  \label{tbl:apgradients}
  \begin{tabular}{lrr}
      \hline
      \multicolumn{1}{c}{Index}&\multicolumn{1}{c}{$d\log I/d\log(r/r_e)$}\\
      \hline
      \hbeta&$0.019\pm0.051$\\
      \mgb&$-0.044\pm0.018$\\
      Fe5270&$-0.049\pm0.018$\\
      Fe5335&$-0.045\pm0.022$\\
      \hline
  \end{tabular}
\end{table}

\begin{table}
  \caption{Stellar population gradients}
  \label{tbl:spgradients}
  \begin{tabular}{lrr}
    \hline
    \multicolumn{1}{c}{Parameter}&\multicolumn{1}{c}{$r_e/4\to r_e$}&
    \multicolumn{1}{c}{$1\farcs35\to r_e$}\\
    \hline
    \logt&$0.01\pm0.05$&$0.01\pm0.05$\\
    \z&$-0.10\pm0.03$&$-0.10\pm0.03$\\
    \enh&$0.00\pm0.01$&$0.00\pm0.01$\\
    \hline
  \end{tabular}
\end{table}

We estimate the mean shifts incurred in the stellar population
parameters when scaling from the LRIS and Moore apertures to `global'
$r=r_e$ apertures.  Unfortunately, we do not have measured effective
radii for many of the galaxies in the Moore sample.  We have however
measured the line strengths and stellar population parameters of the
LRIS galaxies in an equivalently-sized aperture as \citet{M02}.  The
results are presented in Table~\ref{tbl:spgradients}.  As expected
\citep[see, e.g.,][]{T00b}, we find a slight difference in the
metallicity, such that the galaxies are $0.10\pm0.03$ dex more
metal-poor in the `global' $r_e$ aperture than in the smaller
apertures. We find however negligible differences (less than 3 per
cent in each parameter) between the ages and enhancement ratios in the
larger and smaller apertures.

\clearpage

\end{document}